\documentclass{aa}

\usepackage{amsmath,soul,graphicx,txfonts}
\usepackage{xcolor}
\usepackage{mdframed}
\usepackage{enumitem}

\newcommand{\psr}{PSR~B1919+21}

\newcommand{\mbf}[1]{\mbox{\boldmath $#1$}}

\newcommand{\Sec}[1]{\S~\ref{sec:#1}}

\newcommand{\Fig}[1]{Figure~\ref{fig:#1}}
\newcommand{\Figs}[3]{Figures~\ref{fig:#1} #2~\ref{fig:#3}}

\newcommand{\xcovar}{K}
\newcommand{\xcorr}{R}

\renewcommand{\deg}{{\ensuremath{^\circ}}}

\newcommand{\opmswitch}{-5.4\deg}	
\newcommand{\mopmleft}{-7.6\deg}	
\newcommand{\mopmright}{-6.3\deg}	
\newcommand{\ycohright}{-4.7\deg}	
\newcommand{\cmaxleft}{-5.0\deg}	
\newcommand{\cmaxright}{-3.7\deg}	
\newcommand{\torusleft}{-0.8\deg}	
\newcommand{\torusright}{0.8\deg}	
\newcommand{\partorusleft}{-3.6\deg}	
\newcommand{\partorusright}{1.3\deg}	
\newcommand{\saddleleft}{-2.3\deg}	
\newcommand{\saddleright}{-0.3\deg}	

\graphicspath{{./}{figures/}}

\begin{document}

   \title{The polarization of the drifting sub-pulses from PSR~B1919+21}

   \author{N. Primak
          \inst{1}\fnmsep\thanks{primastr@gmail.com},
          C. Tiburzi\inst{2},
          W. van Straten\inst{1},
          J. Dyks\inst{3},
          S. Gulyaev\inst{1}
          }

\institute{Institute for Radio Astronomy \& Space Research,
Auckland University of Technology, Private Bag 92006, Auckland 1142, New Zealand
\and
ASTRON, the Netherlands Institute for Radio Astronomy, Oude Hoogeveensedijk 4, NL-7991 PD, Dwingeloo, the Netherlands
\and
Nicolaus Copernicus Astronomical Center, Polish Academy of Sciences, Rabia\'nska 8, PL-87-100 Toru\'n, Poland}

   \date{Received \today; accepted \today}

  \abstract
   {}
   {We aim to expand our understanding of radio wave emission and propagation in the pulsar magnetosphere by studying the polarization of drifting sub-pulses in highly sensitive observations of PSR~B1919+21 recorded at the Arecibo Observatory.}
   {We apply and compare several methods of analysis and visualization, 
   including eigenvalue analysis of the longitude-resolved covariances between the Stokes parameters; longitude-resolved scatter plots of the normalised Stokes vectors in the Poincar\'e sphere; auto- and cross-correlations between the Stokes parameters as a function of offset in pulse longitude and lag in pulse number; and mean drift bands of polarization state, formed by averaging the Stokes parameters and quantities derived from them synchronously with the drifting sub-pulse modulation period.}
   {We observe regions of pulse longitude where the superposition of orthogonally polarised modes is best described as incoherent and regions
   where the superposition appears to be at least partially coherent.
   Within the region of coherent superposition, over a range of pulse longitudes spanning $\sim 2\deg$, the distribution of the Stokes polarization vectors forms a torus centered near the origin of the Poincar\'e sphere.
   Furthermore, the polarization vectors rotate about the axis of revolution of the torus synchronously with the drifting sub-pulse modulation of the total intensity.}
   {The nearly uniform circular modulation of polarization state, clearly evident in both the toroidal distribution of the Stokes polarization vectors and the mean drift bands of the Stokes parameters, is not predicted by current theoretical models of pulsar emission. We propose different scenarios to explain the generation of the torus,  based on either incoherent or phase-coherent superposition of orthogonally polarised modes.}

   \keywords{polarization -- pulsars: individual: PSR~B1919+21}

\titlerunning{Drifting sub-pulse polarization of PSR~B1919+21}
\authorrunning{N. Primak et al.}

   \maketitle
%

\section{Introduction} \label{sec:intro}

The physics of the pulsar magnetosphere, including the generation and propagation of the observed radio emission, remains poorly understood \citep[e.g.][]{my16,mrm21}.
A viable physical model should explain the origin of coherent emission over a broad range of frequencies, the high degree of polarization, and a wide range of characteristic phenomena, such as nulling \citep[e.g.,][]{bac70}, giant pulses \citep[e.g.,][]{rj01,hkwe03,kbm+06}, the occurrence of orthogonally polarised modes \citep[OPMs; e.g.,]{em69,mth75,scr+84}, and sub-pulse drifting \citep[e.g.][]{dc68}. The models proposed to date are able to explain some of the observed phenomena, but fail to describe others.

Drifting sub-pulses \citep{dc68} are observed in almost half of the pulsar population \citep[e.g.][]{wes06,wse07,bmm19}. 
In addition to the spin period of the pulsar ($P_1$), drifting is usually characterised by two periods: 1) the longitudinal interval or separation between successive drift bands ($P_2$); and 2) the modulation interval, or drift cycle, expressed in pulsar rotations between drift bands ($P_3$). 
Different methods exist to measure these periods. For example, \citet{bac73} and \citet{brc75} introduce the longitude-resolved fluctuation spectrum (LRFS) method, based on one-dimensional Fourier transforms of single-pulse intensities as a function of pulse longitude. 
\citet{dr01} develops the harmonic-resolved fluctuation spectrum, and 
\citet{es02} elaborates an equivalent approach based on the two-dimensional Fourier transform.
\citet[][hereafter, E04]{edw04} extends the LRFS to study the cross-spectral power between the four Stokes parameters, the polarization LRFS (PLRFS).
Drifting sub-pulses are also studied using the longitude-resolved cross-correlation function \citep[e.g.][]{pw86}, two-dimensional auto-correlation function \citep[e.g.][]{tmh75}, and $P_3$ folding \citep[e.g.][]{thhm71,esv03}.

The origin of drifting sub-pulses is unclear. One scenario, known as the ``rotating carousel'' model \citep{rs75,ran93,rrl+06}, suggests that the observed radiation is produced by outflowing streams of charged particles that circulate around the magnetic pole due to \textbf{E} $\times$ \textbf{B} drift, where \textbf{E} and \textbf{B} are the electric and magnetic fields in the vicinity of the polar cap. Owing to the carousel motion, the streams drift across the primary beam of the pulsar, such that the associated sub-pulses appear to shift in pulse longitude as a function of time. 

In relatively rare cases, sub-pulses are observed to periodically switch between orthogonally polarised states synchronously with the drift cycle of the modulated total intensity.
OPM switching is evident as bi-modal distributions of single-pulse position angles (PAs), with peaks offset by $90\deg$ \citep[e.g.][]{em69,mth75,rrs+02}. Many pulsars exhibit OPM switching \citep[e.g.][]{scr+84};
however, to date, only five pulsars are reported to exhibit periodic OPM switching that is synchronised with the drifting sub-pulse modulation: PSRs B0809+74 (\citealt{thhm71}; \citealt{rrs+02}; E04), B1237+25 \citep{rr03}, B0320+39, B0818$-$13 (E04), and B0031$-$07 \citep{iwjc20}.

To explain the synchronicity of the OPM switching and drifting sub-pulses, \citet{rr03} suggest a revision of the carousel model in which the emission is split into two rotating sub-beam structures, each corresponding to one of the observed OPMs.  Owing to birefringence in the pulsar magnetosphere, the orthogonally polarised sub-beams are offset in both azimuthal and meridional directions with respect to the magnetic axis \citep{rrl+06}, such that the observed switching between modes is synchronous with the modulation of total intensity caused by \textbf{E} $\times$ \textbf{B} drifting. 

E04 presents examples of periodic modulations of the Stokes parameters in three pulsars with drifting sub-pulses that cannot be explained by cyclic switching between incoherently superposed OPMs (the result of which would have a linear distribution in Poincar\'e space). Rather, two of the three pulsars exhibit periodic fluctuations such that the tip of the Stokes polarization vector draws an ellipse in Poincar\'e space.
E04 discovers and studies these drift-modulated elliptical variations of the Stokes polarization vector through eigenvalue analysis of the 
PLRFS.  Summing the spectral density tensor of the PLRFS over all harmonics is equivalent to computing the longitude-resolved covariances between the Stokes parameters, which can be used to study stochastic fluctuations of the Stokes parameters \citep{mck04,es04}. \citet{van09} and \citet{vt17} develop the mathematical foundation required to interpret the $4 \times 4$ matrix of covariances between the Stokes parameters by considering three different regimes of incoherent mode superposition (disjoint, superposed, and composite) and the effects of amplitude modulation.

Analysing the covariances between the Stokes parameters provides a way to extend the study of single-pulse variability, such as drifting and OPM switching, to faint pulsars.  For the majority of pulsars, individual pulses cannot be detected and it is not possible to directly study single-pulse variability using scatter plots of the Stokes parameters \citep[e.g.][]{mck04} or longitude-resolved histograms of polarization state \citep[e.g.][]{scr+84,es04}. 
For the typical pulsar, there is also insufficient signal to detect the local sub-pulse drift period $P_3$, which must be estimated in sufficiently short segments of time over which $P_3$ can be considered stable in order to average the Stokes parameters synchronously with the drift period \citep[e.g.][]{thhm71,wes06}.
Therefore, to study the majority of pulsars, it is necessary to develop statistical descriptions of sub-pulse variability that can be used to integrate over a large number of pulses.  These can be applied to develop a more complete picture of single-pulse variability across the entire pulsar population and provide a deeper understanding of the physics of radio wave emission and propagation in the pulsar magnetosphere.

To prepare to study weak sources, we must first study bright pulsars, whose single-pulse variability can be analysed using methods that are applicable only at high signal-to-noise ratio (histograms, scatter plots, $P_3$-folds, etc.) and connected with features derived from fourth-order moments (e.g.\ covariances between the Stokes parameters and the cross-correlations between them as a function of both longitudinal and temporal lags).
These connections contribute toward the development of an interpretive framework from which we can infer the characteristics of single-pulse variability from statistical quantities that can be integrated over sufficiently long intervals.

In this article we present one further step in the development of such a framework through a case study of PSR~B1919+21.
As the first pulsar discovered by Jocelyn Bell \citep{hbp+68}, PSR~B1919+21 is among the most studied and, quite possibly as a consequence of such scrutiny, it has proven more exceptional than exemplary.
For example, \citet{ran83a} first regards its average profile as a ``barely resolved conal-double'', in which the line of sight is neither tangential nor central to a hollow cone of emission.  \citet{rsw89} later argue that PSR~B1919+21 should be classified as a five-component multiple profile, based on evidence that the two main components are each composed of an unresolved pair of components produced as the line of sight crosses inner and outer cones, and that the saddle region between the two main components exhibits properties of core emission. With the caveat that there appears to be a weak trace of central core emission, \cite{omr19} classify PSR~B1919+21 as having a conal quadruple profile produced by a line of sight that that cuts both inner and outer cones more centrally without passing near the core beam.  
With a spin-down energy of $\dot E \sim 2.2 \times 10^{31}$, the
radio signal from PSR~B1919+21 is expected to be dominated by conal emission \citep{row20}.
Core emission that supports a five-component multiple profile classification appears to be more evident at radio frequencies below 80~MHz \citep{bgt+21}.

Compounding the challenges associated with classifying its emission, PSR~B1919+21 shows no evidence of the radius-to-frequency mapping thought to be associated with outer-cone component pairs \citep{mr02a}.
Its main pulse is also preceded by weak emission \citep{wcl+99} that appears more clearly at high radio frequencies as a separate leading component well beyond its putative outer cone \citep{omr19}.
Furthermore, it is not possible to reliably constrain the magnetic field and viewing geometry of PSR~B1919+21, owing to complex variations of its longitude-resolved position angle (P.A.) that depart significantly from the predictions of the rotating vector model \cite[RVM;][]{rc69a}.  
For many pulsars, deviations from the RVM can be described by transitions in the dominance of orthogonally polarised modes of emission, most clearly evident as jumps of 90$\deg$ in the P.A.\ curve \citep[e.g.][]{ew01}.  However, at some radio frequencies, PSR~B1919+21 exhibits a sharp 45$\deg$ jump in P.A.\ near the
profile peak, which can be interpreted as a sudden narrowing of the distribution of phase delay between the natural modes of wave propagation in the pulsar magnetosphere \citep{dyk19}.
In Fig.~18 of \citet{mar15}, some of the smaller distortions of the P.A. profile appear to be associated with absorption features in the total intensity that were first detected at frequencies below 200 MHz \citep{han73,cor75}.
These striking features of both the single-pulse and average profiles of PSR~B1919+21 remain poorly understood.

Although the first-discovered pulsar clearly warrants further study, none of the above difficulties or exceptions motivate its selection for this work.  Rather, as described in \Sec{data}, PSR~B1919+21 was objectively selected owing to the anomalous distribution of its polarization state, as identified by applying novel statistical methods to publicly available data.
In \Sec{meth_res} we report on the results of applying a wide variety of methods to analyse the variability of its single-pulse polarization state. 
In \Sec{discussion} we discuss and interpret the results, which are summarised in \Sec{conclusions}.

\section{Data set} \label{sec:data}

The observations used in this article are part of a larger single-pulse dataset collected at the Arecibo Radio Telescope, mainly from 1988 to 1992, at multiple frequencies, and presented and described in detail by \citet{hr10}. From the polarization-calibrated full Stokes data summarised in Table 2 of \citet{hr10}, we selected 32 pulsars observed at one or two central frequencies, 1414 MHz and 430 MHz, as part of the P1260 program.\footnote{These data are located at \url{http://www.uvm.edu/~pulsar/P1260}.}

After computing the eigenvalues of the longitude-resolved covariances between the Stokes parameters for each observation,
we searched for behaviour that is inconsistent with incoherent OPM superposition \citep{vt17}.
That is, we searched for evidence of correlations between the Stokes parameters that are not consistent with a prolate spheroidal distribution of the polarization vector.
The observation of PSR~B1919+21 at 1414~MHz clearly stands out as anomalous because, over a wide region of pulse longitude, the distribution of the Stokes polarization vector is characterised by an oblate spheroid. 

PSR~B1919+21 produces complex patterns
in the P.A.\ distributions of single pulse polarization state \citep[e.g.\ Fig. 18 of][]{mar15}.
It also exhibits sub-pulse drift with band separations $P_2 \approx 10 - 17$ ms and $P_3 \approx 4 - 4.4$ $P_1$ \citep[e.g.,][]{bac73,han73,cor75,wol78,pw86,wse07}. 
%
%
%
It has three distinct drifting components that are not well separated in pulse longitude; that is, the sub-pulses from different components drift through shared regions of pulse longitude.
The definitions of the drift-band region boundaries depend on radio frequency; however, $P_3$ is not frequency-dependent and is common to all three drift band regions \citep{pw86}.
Although the single-pulse polarization and sub-pulse drifting properties of PSR~B1919+21 have been separately analysed, the connection between them has not been studied in detail.\footnote{\citet{mth75} noted the possibility of correlation between polarization state and the drifting sub-pulse structures.}
Therefore, the remainder of our analysis focuses on PSR~B1919+21.


\section{Methods and results} \label{sec:meth_res}

The statistical methods applied in this work begin with an ensemble of sample mean Stokes 4-vectors
\begin{equation}
\mathrm{S}(\phi,t) = [\mathrm{I}(\phi,t),\mathrm{Q}(\phi,t),
\mathrm{U}(\phi,t),\mathrm{V}(\phi,t)]^T
\end{equation}
sampled as a function of pulse longitude $\phi$, and pulsar rotation (or turn) $t$. Here, $x^T$ denotes the transpose of vector $x$; i.e., the Stokes 4-vector is treated as a column vector.
The sample mean Stokes parameters are averaged over a single interval of pulse longitude, which for the PSR~B1919+21 data presented in this study is $\sim 0.325\deg$, or about 1.2~ms.
The $4\times4$ matrix of cross-covariances between the Stokes parameters at two pulse longitudes, $\phi_1$ and $\phi_2$, as a function of time-lag in pulsar rotations $\tau > 0$ is given by
\begin{equation} 
     \xcovar(\phi_1, \phi_2; \tau) = \frac{1}{N_\mathrm{pulse}}
     \sum_{t=\tau+1}^{N_\mathrm{pulse}} \bar{S}(\phi_1,t-\tau)  \otimes \bar{S}(\phi_2,t)^T,
     \label{eqn:xcovar}
\end{equation}
where $N_\mathrm{pulse}$ is the number of pulses, $\bar{S}$ represents the Stokes parameters after subtracting a local mean value, and $\otimes$ represents an outer product.
If $\xcovar_{\mu\nu}(\phi_1,\phi_2;\tau)$ is the element in row $\mu$ and column $\nu$, 
then the cross-covariance at negative lag is given by transposing both $\mu$ and $\nu$, and 
$\phi_1$ and $\phi_2$; i.e.
\begin{equation} 
\xcovar_{\mu\nu}(\phi_1, \phi_2; -\tau) = \xcovar_{\nu\mu}(\phi_2, \phi_1; \tau).
\end{equation}

\subsection{ Longitude-resolved covariances between the Stokes parameters}  \label{sec:eigen}

As in \citet{mck04} and \citet{es04}, we first studied the longitude-resolved $4\times4$ matrix of covariances between the Stokes parameters,
\begin{equation} 
    C(\phi) = \xcovar(\phi, \phi; 0)
    \label{eqn:phaseRes_covar_matrix}
\end{equation}
by computing the eigendecomposition of the $3\times3$ partition of $C$ corresponding to the three-dimensional polarization vector, $\mbf{S}=(\mathrm{Q, U, V})$. The eigenvectors and eigenvalues derived from this analysis describe the distribution of the polarization vectors in Poincar\'e space. At a given pulse longitude $\phi$, the eigenvectors define the principle axes of the distribution of the polarization vectors; the eigenvalues $\lambda_i$, $i\in\{1,2,3\}$, are equal to the variances of the distribution in the directions of the corresponding eigenvectors. For example, when $\lambda_{1} > \lambda_{2} = \lambda_{3}$, the distribution of the polarization vectors has a prolate spheroidal shape, as expected for incoherent OPM superposition \citep{vt17}.

\begin{figure}
\centering
\centerline{\includegraphics[width=86mm]{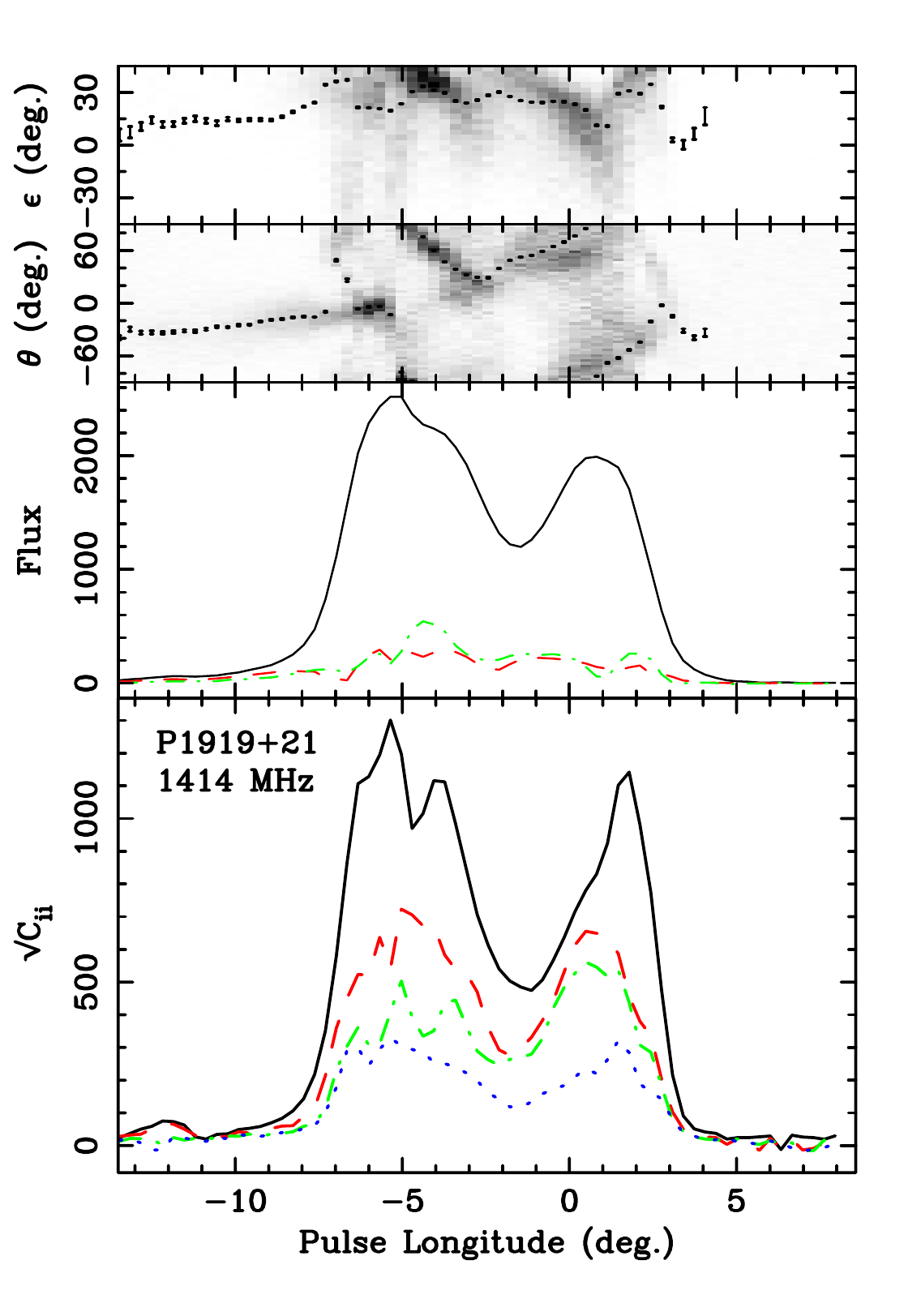}}
\caption{ Longitude-resolved mean and variability of the polarised emission from PSR~B1919+21 at 1414~MHz. Panel a) shows both the ellipticity angle of the average Stokes parameters (points with error bars) and a histogram of the ellipticity angles of the individual pulses (greyscale intensity). Similarly, panel b) shows both the position angle of the average Stokes parameters and a histogram of the position angles of the individual pulses. Panel c) shows the integrated profile in total intensity (black, upper solid), linear (red, dashed) and circular polarization (blue, dot-dashed). These profiles were obtained by averaging together all the available data. In panel d), the black solid line shows the standard deviation of the total intensity, and the red dashed, green dot-dashed, and blue dotted lines display the square roots of the eigenvalues (from largest to smallest, respectively) of the 3$\times$3 partition of the covariance matrix that describes the fluctuations of the polarization vector $\mbf{S}=(Q,U,V)$. }
\label{fig:stat_verif_1919_1}
\end{figure}

The derived eigenvalues are shown in panel d) of \Fig{stat_verif_1919_1}. 
%
%
Over the region of pulse longitude $-5\deg \lesssim \phi \lesssim 3\deg$,
the eigenvalues indicate an unexpected and uncommon \textit{oblate distribution} of the polarization state, which is the primary focus of the following section.
The longitude-resolved standard deviation of the total intensity also highlights a component around longitude $-12\deg$ that is most likely related to the precursor detected at 4.6 GHz \citep{omr19}.  The phase-resolved modulation index of this component is $\sim$ 1.2 to 1.3, around two to three times the modulation index in the main profile, which varies between $\sim$ 0.4 and 0.7. 

\subsection{Distribution of the polarization vectors}
\label{sec:Psphere}

\begin{figure}
\centering
\includegraphics[scale=0.25]{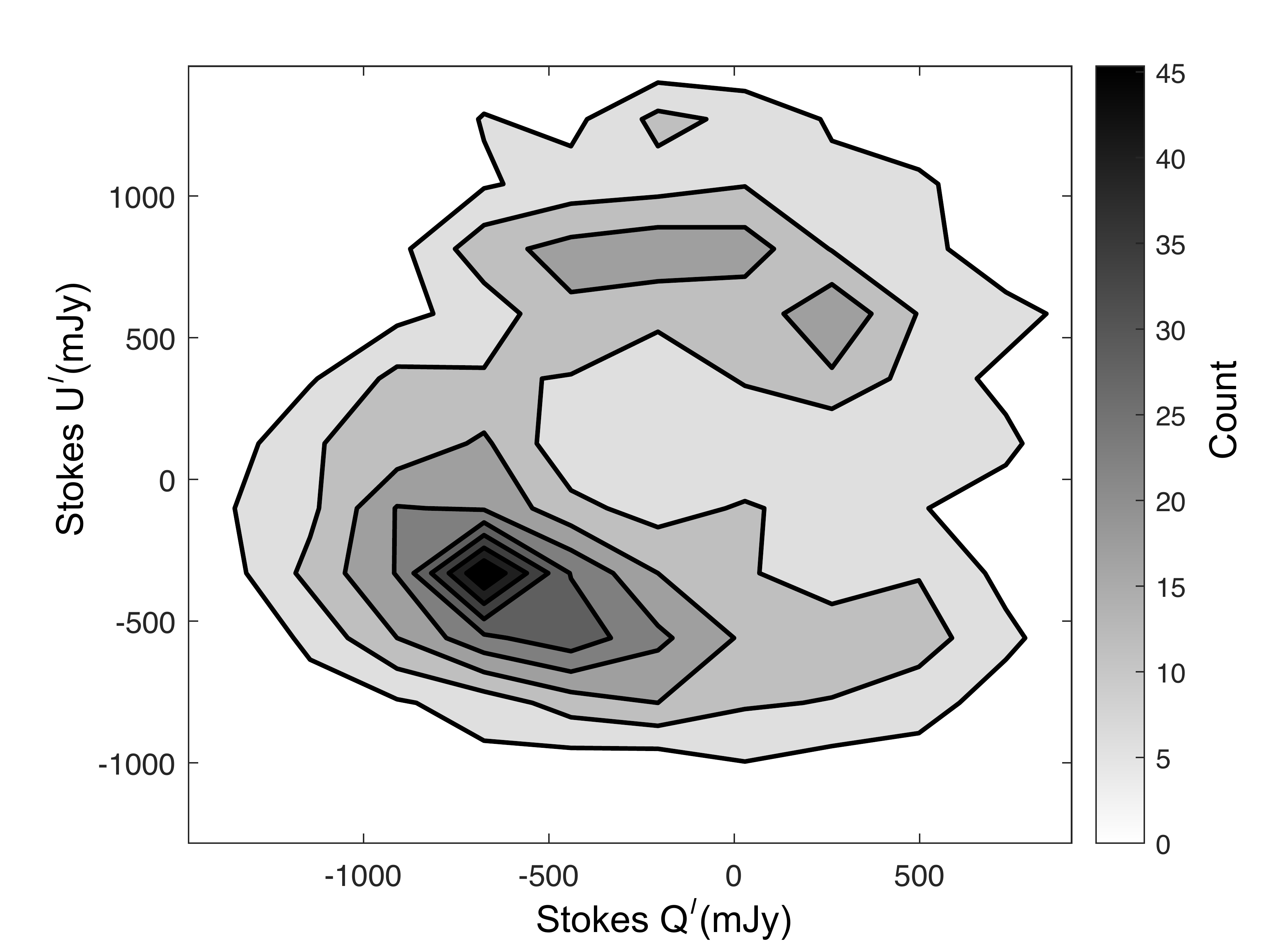}
\caption{Integrated number of single-pulse polarization states projected onto a two-dimensional plane that is normal to the minor axis of the oblate spheroidal distribution observed at pulse longitude 0$\deg$. The greyscale intensity represents the count of instances in each pixel after integrating 1100 single pulses; contours are spaced at intervals of 5 counts. 
The angle between the Stokes V axis and the normal to the Q'--U' plane is $7.2\deg$.}
 
\label{fig:donut_no_norm}
\end{figure}

As in \citet{mck04}, we produced scatter plots of the Stokes polarization vector using two-dimensional projections of the 3-dimensional Poincar\'e space. 
In \Fig{donut_no_norm}, one such projected scatter plot is divided into a $25\times25$ grid and the number of points in each pixel is integrated.
Rather than the expected concentration of points near the origin, the polarization states are concentrated on the circumference of a ring that is offset from the origin.
%
%

If the Stokes vectors are normalised by the total intensity, then the normalised polarization vectors will lie within a sphere of unit radius known as the Poincar\'e sphere. 
Visualizing the polarization state in the Poincar\'e sphere offers a couple of advantages: first, the length of the normalised polarization vector indicates the degree of polarization; second, assuming that all four Stokes parameters are modulated by the same function, normalization by Stokes I reduces additional scatter due to amplitude-modulation that is intrinsic to the pulsar and/or owing to scintillation in the interstellar medium.

\begin{figure}
\centering
\includegraphics[scale=0.4]{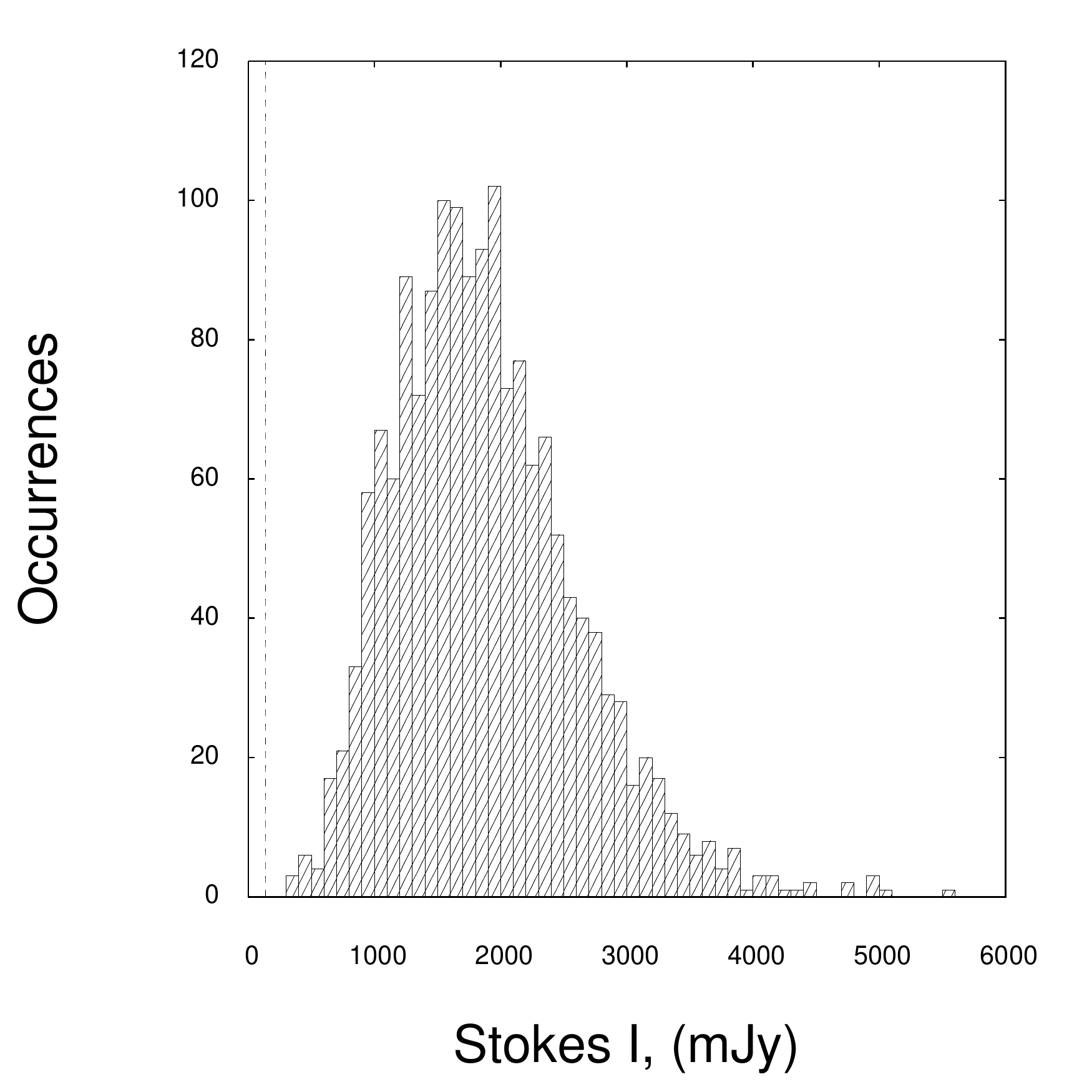}
\caption{Total intensity histogram at pulse longitude 0$\deg$, where the Stokes parameters are toroidally distributed. This example demonstrates the validity of Stokes parameters normalization.  Stokes I has a similar behaviour throughout all pulse longitudes. The vertical dashed line shows the maximum value of the median off-pulse noise. }
\label{fig:i-hist}
\end{figure}

\begin{figure}
\centering
\includegraphics[width=92mm,trim=42mm 0 30mm -3mm,clip]{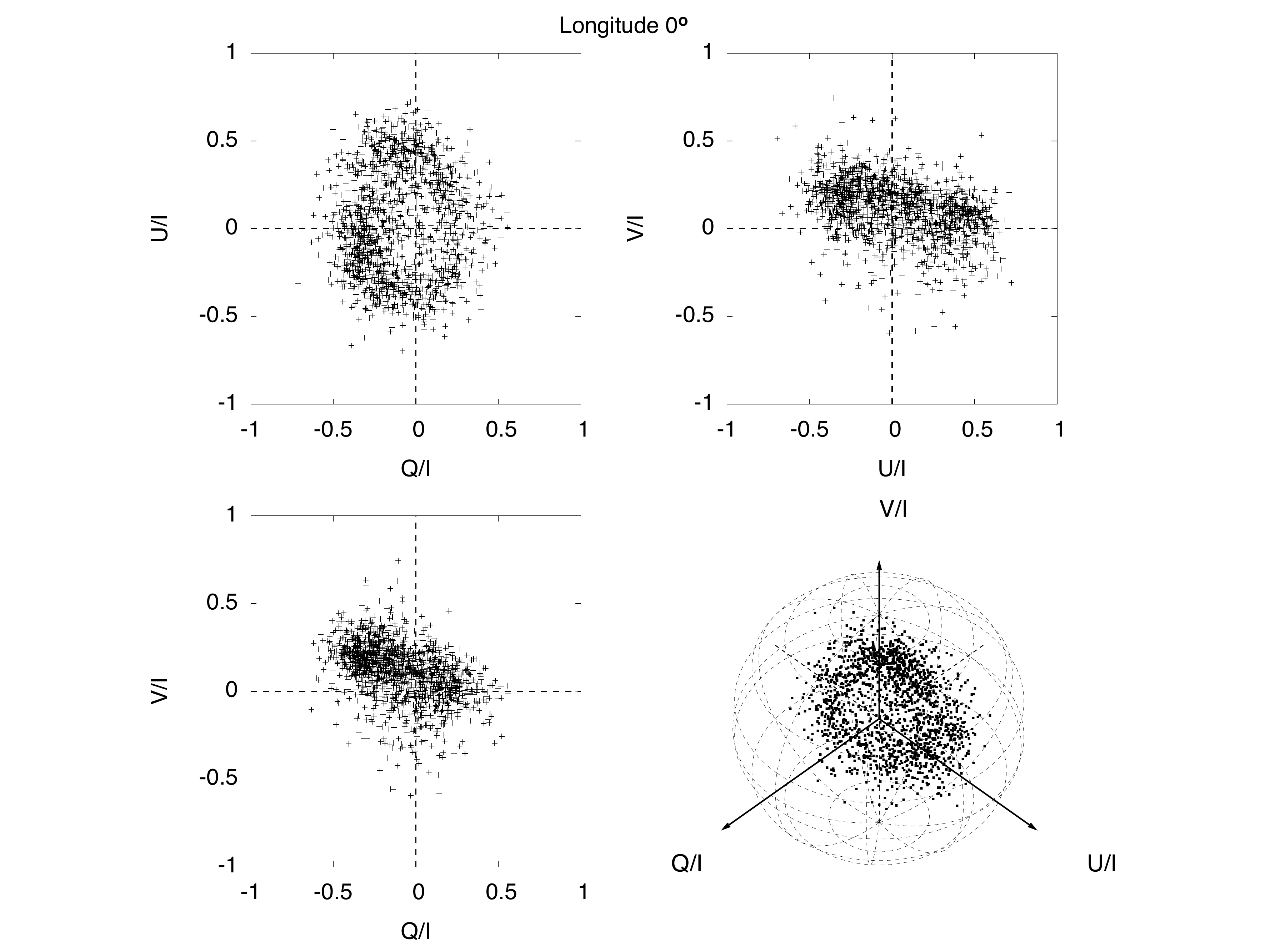}
\caption{Distribution of the normalised Stokes parameters in the Poincar\'e sphere at $\phi$ 0.$\deg$. The two upper panels and the lower-left panel show the projected, 2D distribution of the Stokes parameters in the three planes normal to the basis vectors, and the 3D distribution is displayed in the lower-right panel.}
\label{fig:torus_1919_1}
\end{figure}

Normalization by Stokes I is possible only for very bright single-pulse observations, where the total intensity is significantly different from zero. 
Therefore, to verify that this normalization is valid, we plot histograms of Stokes I in the on-pulse regions of interest. As an example, \Fig{i-hist} shows that the total intensity is well above zero at $\phi=0\deg$, where the oblate distribution of the Stokes vectors is observed. 
The median of the off-pulse noise (i.e., the median value of the standard deviation estimated from the off-pulse longitudes of 200 consecutive pulses) varies as a function of time, most likely owing to the source moving away from the principal axis of the primary reflector.  At the start of the observation, the median off-pulse noise is 75 mJy and at the end it was 134 mJy.  The vertical dashed line in \Fig{i-hist} depicts the maximum value of the standard deviation of the off-pulse noise (at the end of the observation).
%
%
This figure shows that, at this pulse longitude, it is possible to visualize scatter plots of the polarization state in either Poincar\'e space or within the Poincar\'e sphere.
In \Fig{torus_1919_1}, the projected scatter plots of the normalised polarization vectors at $\phi=0\deg$ show that the distribution is not an oblate spheroid, but a torus of revolution.
%
A toroidal distribution is clearly detectable at  $\torusleft\lesssim\phi\lesssim\torusright$, 
and a partial torus appears to be visible in adjacent pulse ranges $-3.9\deg \lesssim \phi \lesssim -1\deg$, and $1\deg \lesssim \phi \lesssim 1.6\deg$. 
Scatter plots for four neighboring longitudes are presented in \Fig{94_torus_1919_1}; these show that, as a function of pulse longitude, the torus changes orientation and shifts its center. Moreover, the centers of the toroidal distributions are offset from the origin of Poincar\'e sphere axis at all longitudes.
For comparison, we also present projected scatter plots of the normalised Stokes polarization vector observed at $\phi\simeq-6.8\deg$ in \Fig{75_torus_1919_1}. The  distribution shown in this scatter plot is consistent with Figure~\ref{fig:stat_verif_1919_1}, where the eigenvalues at this longitude indicate a prolate spheroid.

\subsection{Correlations of Stokes parameters}  \label{sec:correlations}

Together with the 3D distributions of the Stokes parameters it is important to know whether the torus formed in an orderly or a random way. To this end, we explored the spatial and temporal relationships between the polarization vectors via

\begin{enumerate}[label=(\roman*)]
\item the longitude-offset cross-correlation function of the Stokes parameters with respect to a reference longitude; and 
\item the longitude-resolved cross-correlation functions between two different Stokes parameters.
\end{enumerate}

For each element of $\xcorr(\phi_1, \phi_2; \tau)$, we computed the Pearson correlation coefficient,
\begin{equation} 
    \xcorr_{\mu\nu}(\phi_1, \phi_2; \tau) = 
    \frac{\xcovar_{\mu\nu}(\phi_1, \phi_2; \tau)}
    {\sigma_\mu(\phi_1)\sigma_\nu(\phi_2)}
    \label{eqn:pearson_coef}
\end{equation}
where $\mu,\nu\in\{$I,Q,U,V$\}$, and the longitude-resolved variance is given by
\begin{equation} 
    \sigma^2_\mu(\phi) = C_{\mu\mu}(\phi)
                       = \xcovar_{\mu\mu}(\phi, \phi; 0)
\label{eqn:sigma}
\end{equation}

Prior to computing $\xcorr(\phi_1, \phi_2; \tau)$, we subtracted the moving average of the Stokes parameters from each pulse longitude, computed with a boxcar size 100 pulse periods, which is smaller than the scintillation time
at 1.4~GHz \citep[$\sim 500$ to 1500~s;][]{cor86,brg99}.

\subsubsection{ Longitude-offset cross-correlations for each Stokes parameter} 

\begin{figure*}
\centering
\includegraphics[width=150mm]{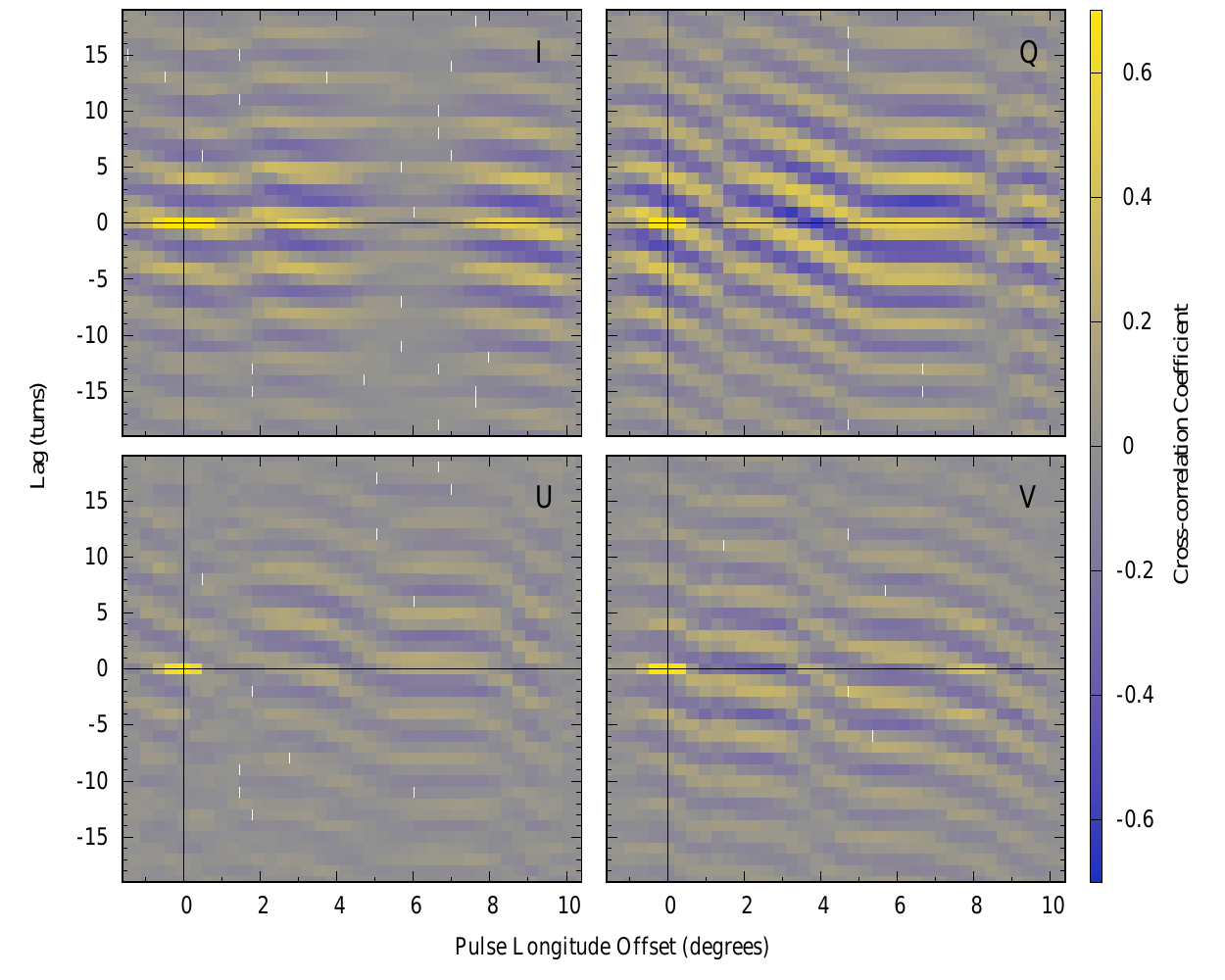} 
\caption{Longitude-offset cross-correlations of each Stokes parameter
as a function of lag $\tau$ and longitude offset $\Delta\phi$ with respect to the reference pulse longitude $\phi_0 \simeq -6.8\deg$.
In each panel, a horizontal black line marks zero lag and a vertical black line marks zero longitude offset.}
\label{fig:2DAC}
\end{figure*}

\Figs{2DAC}{and}{2DAC_b} present the longitude-offset cross-correlation coefficients,
\[
R^\prime_\mu(\Delta\phi,\tau;\phi_0)
= \xcorr_{\mu\mu}(\phi_0, \phi_0+\Delta\phi; \tau),
\]
with $\mu\in\{$I,Q,U,V$\}$, as a function of lag $\tau$ and longitude offset $\Delta\phi$ with respect to the reference pulse longitude $\phi_0$.

\begin{figure*}
\centering
\includegraphics[width=150mm]{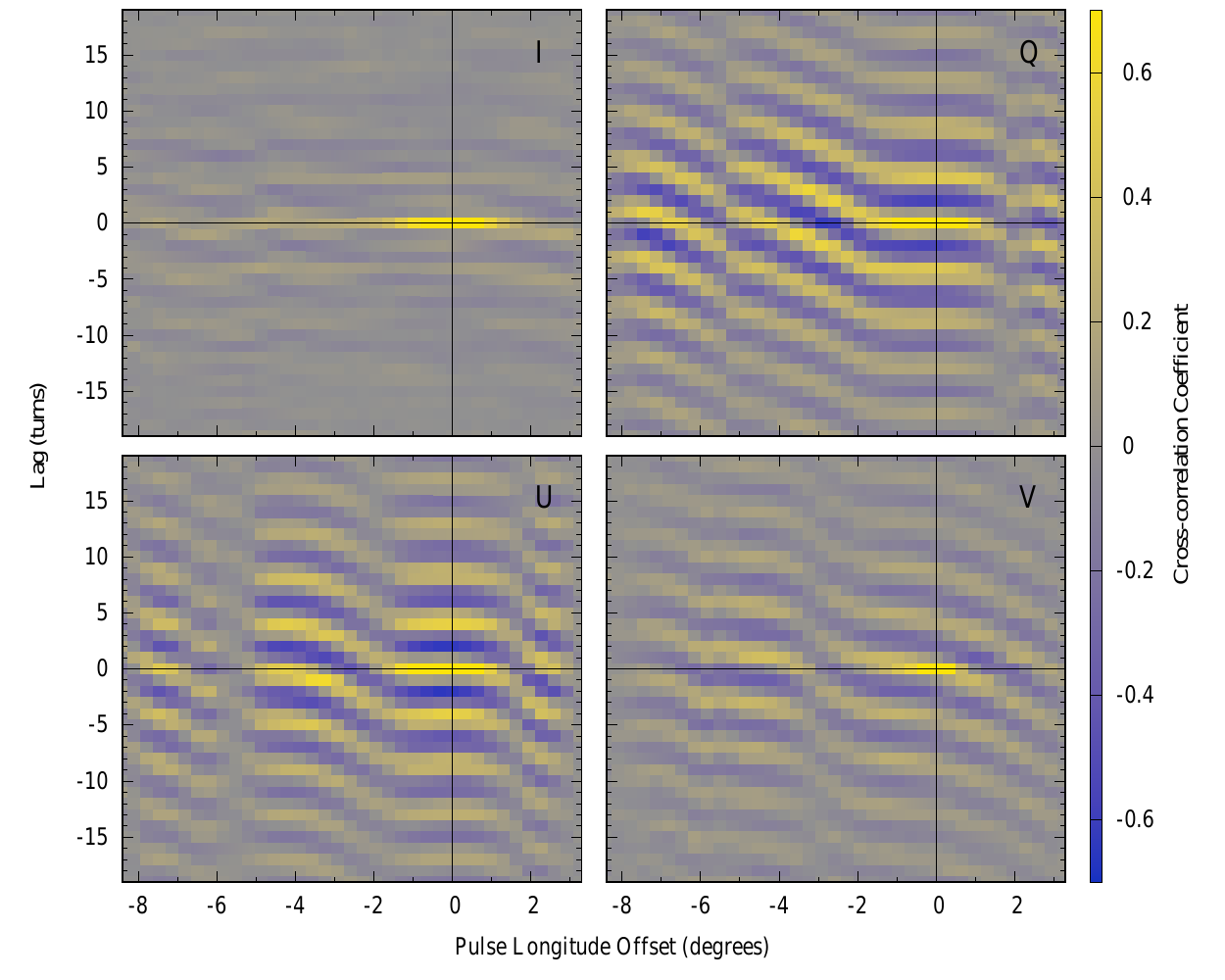}
\caption{As for Figure~\ref{fig:2DAC}, but for reference longitude $\phi_0=0\deg$, in the middle of the region where the toroidal distribution of the polarization vector in the Poincar\'e sphere is observed.}
\label{fig:2DAC_b}
\end{figure*}

The reference longitude of \Fig{2DAC} is $\phi_0 \simeq -6.8 \deg$, which is within a region of pulse longitude that is characterised by a prolate spheroidal distribution of the Stokes parameters that is approximately centered at the origin of the Poincar\'e sphere. 
\Fig{2DAC} shows three main drift regions in total intensity, as first noted by \citet{pw86}.  In this plot, the first drift-band lies between the beginning of the on-pulse region and a pulse longitude of approximately $\opmswitch$ (corresponding to a longitude offset of $1.4\deg$), the second drift band starts at the end of the first band and spans up to a longitude of about $-2.6\deg$ (longitude offset $4.5\deg$), and the third drift band lies approximately between longitudes $-0.6\deg$ and $3\deg$.

The periodic modulation of total intensity at the reference longitude
is remarkably correlated with the modulation in the rest of the on-pulse region, even as far away as longitude offset $\Delta\phi \simeq 9.7 \deg$.
However, this correlation is diminished over the saddle region $5\deg \lesssim \Delta\phi \lesssim 7\deg$ (corresponding to longitudes $-1.8\deg\lesssim\phi\lesssim 0.2\deg$). This region of diminished correlation in total intensity partially overlaps the region ($5.5\deg \lesssim\Delta\phi \lesssim7.5\deg$) where the periodic modulations of Stokes Q and U exhibit little or no drifting (i.e., the slope is flat). The periodic modulations of Stokes Q, U, and V remain strong in this region.

The reference longitude of \Fig{2DAC_b} is $\phi_0 = 0 \deg$, which is in the center of the longitude region where the toroidal distribution of Stokes (Q, U, V) in the Poincar\'e sphere is observed. \Fig{2DAC_b} reaffirms that the total intensity in this region is not highly correlated with intensity in the rest of the on-pulse region. 
In both \Fig{2DAC} and \Fig{2DAC_b}, the longitudinal discontinuity in the modulation pattern of Stokes Q
at $\phi\simeq\opmswitch$ occurs at the transition between OPMs that is also evident in \Fig{stat_verif_1919_1}.

\subsubsection{ Longitude-resolved Stokes Q,U,V cross-correlations}

\begin{figure}
\centerline{\includegraphics[width=82mm]{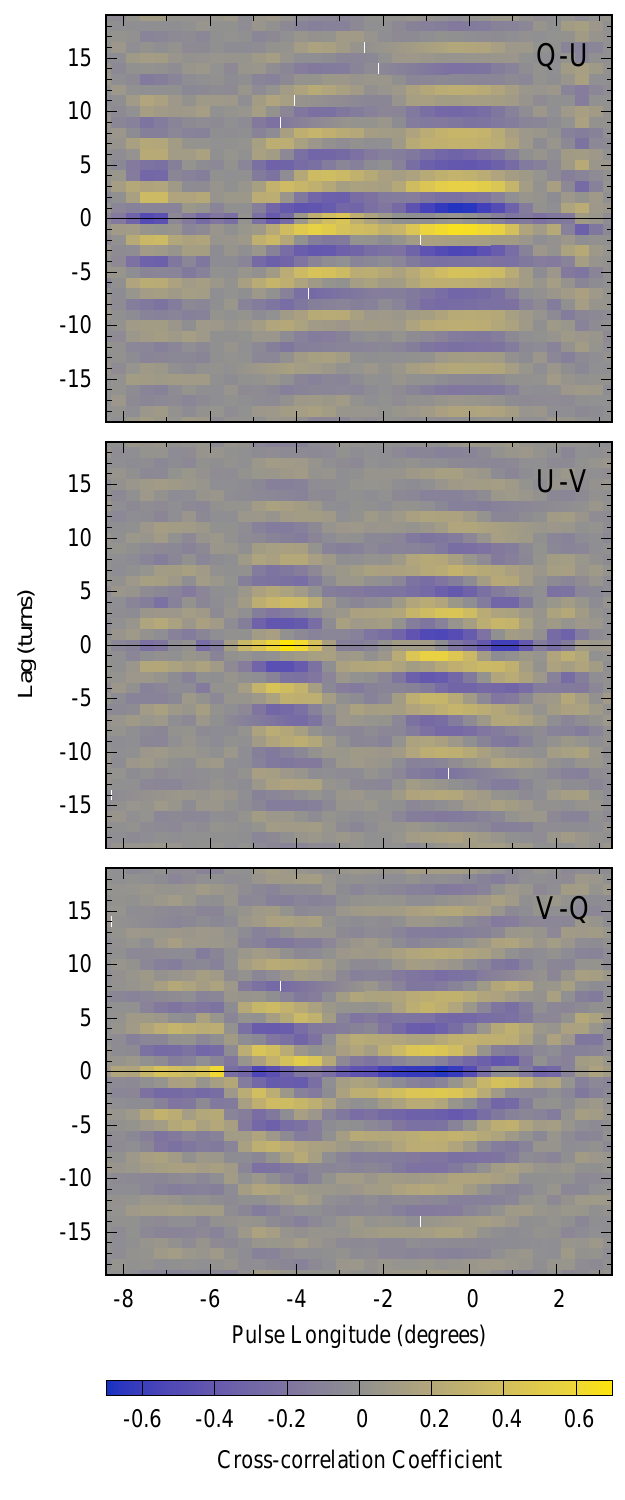}}
\caption{Longitude-resolved Stokes Q,U,V cross-correlations as a function of lag $\tau$ and pulse longitude $\phi$.  From top to bottom are
plotted the correlations between Stokes Q and U, Stokes U and V, and Stokes V and Q. In each panel, a horizontal black line marks zero lag.}
\label{fig:XcorrPhBin}
\end{figure}

\Fig{XcorrPhBin} presents the longitude-resolved cross-correlation coefficients,
\[
R^\prime_{\mu\nu} (\phi,\tau)
= \xcorr_{\mu\nu}(\phi, \phi; \tau),
\]
with $(\mu,\nu)\in\{$(Q,U),(U,V),(V,Q)$\}$, as a function of lag $\tau$ and pulse longitude $\phi$.
Owing to symmetry, these three panels depict all of the information contained in the 6 off-diagonal components of the $4 \times 4$ matrix, $R^\prime(\phi,\tau)$; furthermore, taking the one-dimensional Fourier transforms of the components of $R^\prime(\phi,\tau)$ along the $\tau$ axis yields the PLRFS studied by E04.
As in the PLRFS, the correlations between the Stokes parameters indicate their cyclic oscillation about an elliptical path in Poincar\'e space.  They also indicate that the normal to this elliptical path precesses as a function of pulse longitude.

For example, over $-1.5\deg \lesssim\phi \lesssim 1\deg$, $R^\prime_{\mathrm{QU}}(\phi,\tau)$ is antisymmetric about $\tau = 0$, 
which indicates that the oscillations of Stokes Q and U are 90\deg out of phase.
This is consistent with the periodic cycling of the polarization vector around an elliptical path in a plane that is only slightly offset from the Q--U plane\footnote{The offset varies from pulse longitude to pulse longitude.} over this range of pulse longitudes.

At earlier pulse longitudes within this region, e.g.\ $-1.5\deg \lesssim \phi \lesssim 0\deg$,
$R^\prime_{\mathrm{VQ}}(\phi,\tau)$ is negative and symmetric about $\tau = 0$, which indicates that Stokes Q and V are anti-correlated ($180\deg$ out of phase). 
This is consistent with the anti-correlation that is visible in the Q--V scatter plots in \Fig{torus_1919_1} and the upper panels of Figure~\ref{fig:94_torus_1919_1}.  
Furthermore, the oscillations of Stokes U and V are 90\deg out of phase, as indicated by the antisymmetry of $R^\prime_{\mathrm{UV}}(\phi,\tau)$. 
Both the anticorrelation of Stokes Q and V and the antisymmetry of Stokes U and V are consistent with an axis of revolution of the torus that lies primarily in the Q-V plane, such that there is an apparently elliptical path projected into the U-V plane.

Similarly, at later pulse longitudes within this region, e.g.\ $0\deg \lesssim \phi \lesssim 1\deg$,
$R^\prime_{\mathrm{UV}}(\phi,\tau)$ is negative and symmetric about $\tau = 0$, which indicates that Stokes U and V are anti-correlated. 
This is consistent with the anti-correlation that is visible in the U--V scatter plots in \Fig{torus_1919_1} and the lower panels of Figure~\ref{fig:94_torus_1919_1}.  
Furthermore, the oscillations of Stokes Q and V are 90\deg out of phase, as indicated by the antisymmetry of $R^\prime_{\mathrm{VQ}}(\phi,\tau)$. 
Both the anticorrelation of Stokes U and V and the antisymmetry of Stokes Q and V are consistent with an axis of revolution of the torus that lies primarily in the U-V plane, such that there is an apparently elliptical path projected into the Q-V plane.
That is, as a function of pulse longitude, the axis of revolution of the torus precesses about the Stokes V axis, rotating from primarily in the Q--V plane to primarily in the U-V plane.

\subsection{Average drift band}

\begin{figure}
\centering
\includegraphics[trim=34mm 20mm 30mm 5mm, width=92mm]{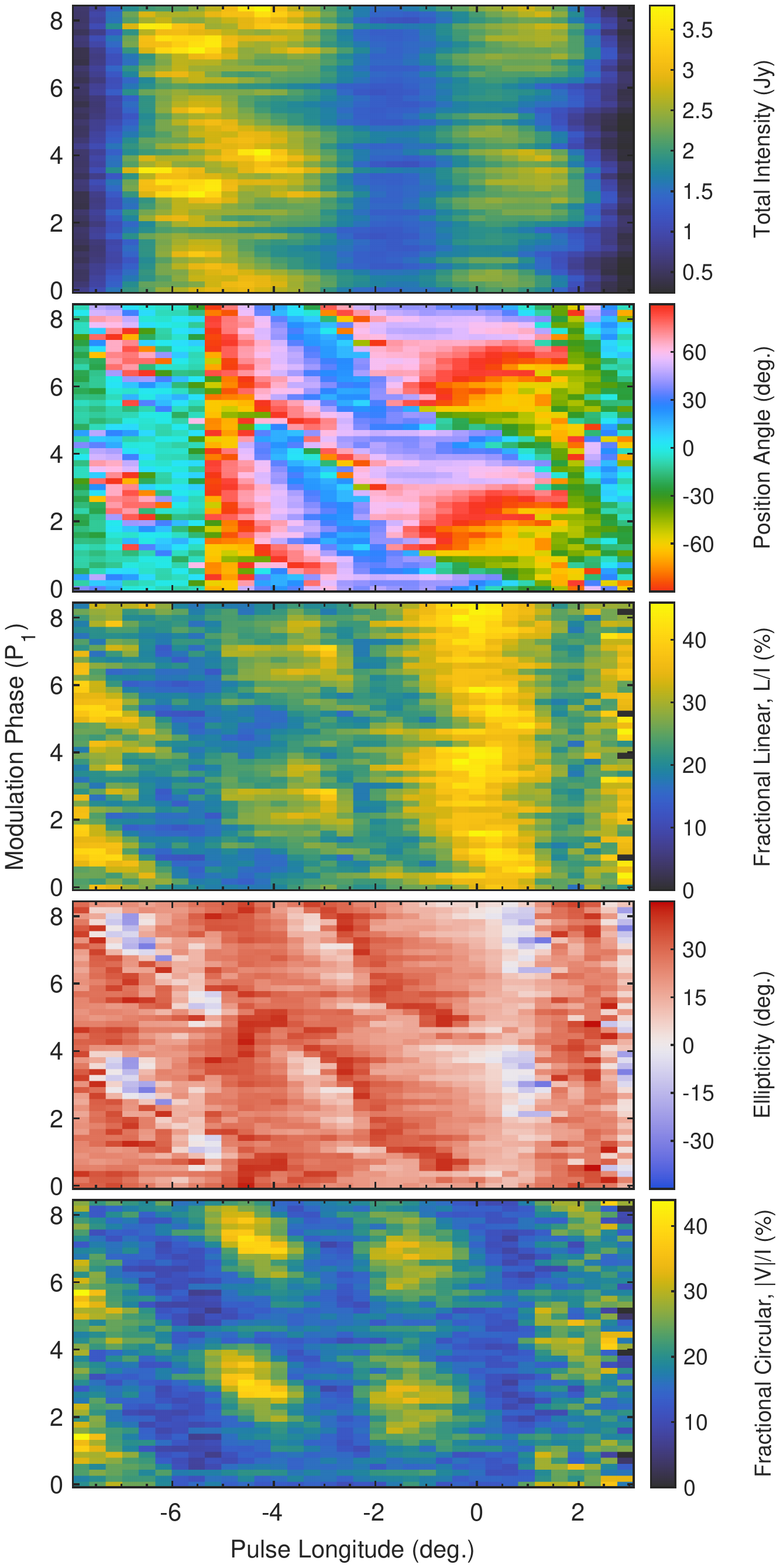}
\caption{$P_3$-folded polarization state of the first 500 pulses in the  PSR~B1919+21 dataset. From top to bottom are the total intensity; the position angle, $\theta=\tan^{-1}(U^\prime/Q^\prime)/2$; the fractional linear polarization, $\hat{L}=\sqrt{Q^2+U^2}/I$; the ellipticity angle, $\epsilon=\sin^{-1}(V^\prime/L^\prime)/2$; and the fractional circular polarization, $\hat{C}=|V|/I$; where $P_3$-folded quantities are shown with primed symbols (e.g.\ $Q^\prime$ and $U^\prime$) and the symbols for single-pulse quantities are not primed (e.g.\ $Q$ and $U$).
That is, $\hat{L}$ and $\hat{C}$ are derived from the single-pulse Stokes parameters before $P_3$ folding and $\theta$ and $\epsilon$ are derived from the $P_3$-folded Stokes parameters.
Furthermore, $\hat{L}$ and $\hat{C}$ are integrated only when the single-pulse total intensity exceeds 4 times the standard deviation of the of noise.
In each panel, the modulation period ($P_3 \simeq 4.2$ $P_1$) is plotted twice to better illustrate the continuity of the pattern.}
\label{fig:P3fold}
\end{figure}

As in \citet{thhm71} and \citet{esv03}, we estimated the polarization state of
the mean drift bands by averaging the single-pulse Stokes parameters and quantities derived from them synchronously with the modulation period, $P_3 \simeq 4.2 P_1$, which was determined via short-time Fourier transform analysis.  
The results of $P_3$ folding the first 500 pulses in the data set are shown in \Fig{P3fold}, which exhibits several noteworthy features.
%

In all panels, pulse longitude $\phi\simeq\opmswitch$ marks several significant changes in the sub-pulse modulation pattern. First, there is a $180\deg$ phase shift in the modulation of the total intensity associated with the transition from region 1 to region 2
of the mean drift band. 
 The overlap of drift regions 1 and 2 leads to an apparent doubling of the modulation frequency, which is also detected in Fig.~A.21 of \citet{wse07}, where it is described as ``$P_3 = 2 P_0$ flickering''. 
Overlapping and offset drift regions also explain the reduced the longitude-resolved modulation index observed
over this region of pulse longitude \citep{wes06}.

The transition from drift region 1 to drift region 2 is accompanied by a longitudinal transition between orthogonally polarised modes, which is also clearly visible in the position angle histogram plotted in panel b) of \Fig{stat_verif_1919_1}. Before this transition (over $\mopmright\lesssim\phi\lesssim\opmswitch$), the position angle exhibits little modulation and has a mean value close to $0\deg$.  Immediately after this transition (over  $\opmswitch\lesssim\phi\lesssim\ycohright$), the position angle varies between $\sim -60\deg$ and $\sim 75\deg$ as a function of modulation phase.
The minimum degrees of polarization (both fractional linear and fractional circular) in \Fig{P3fold} occur just before $\phi\simeq\opmswitch$.  Shortly after this transition (over $\cmaxleft\lesssim\phi\lesssim\cmaxright$) the ellipticity angle exhibits minimal modulation and remains greater than $15\deg$; correspondingly, the fractional circular polarization also reaches its maximum value in this pulse longitude range.

In the longitudes shortly before the OPM switch at $\phi\simeq\opmswitch$ (between approximately $\mopmleft$ and $\mopmright$) the $P_3$-folded position angle and ellipticity angle indicate variation between orthogonally polarised states as a function of modulation phase.  However, as first noted for PSR~B0809+74 \citep{thhm71}, the position angle does not jump discontinuously from one mode to the other, as expected for incoherent OPM superposition; rather, there is a continuous decrease followed by a continuous increase in position angle as a function of modulation phase.  With reference to Figure~\ref{fig:75_torus_1919_1}, the apparent continuous oscillation of the polarization state could be explained by periodic switching along an axis that does not pass through the origin in the Q--U plane.  A likely origin for such off-center switching is the incoherent transition between slightly non-orthogonal polarization modes \citep[e.g.][]{mck03a}.
Similarly, continuous oscillation of position angle as a function of modulation phase is observed in the longitude range following the OPM transition, between approximately $\opmswitch$ and $\partorusleft$. Here, there is also a clear systematic swing of position angle through each sub-pulse as a function of longitude and, as first noted for PSR~B0809+74 \citep{thhm71}, this swing follows the sub-pulse structure as it drifts in longitude as a function of modulation phase.  This swing gives rise to the continuous oscillation of position angle at a fixed longitude as a function of modulation phase.

Over the pulse longitudes that exhibit the toroidal and partially toroidal distributions (approximately $\partorusleft \lesssim \phi \lesssim \partorusright$) the position angle no longer oscillates back and forth; rather, it cycles monotonically through the full spectrum of visible colours, clearly indicating the elliptical motion of the polarization vector in the Q--U plane.  Here, the degree of linear polarization is also a maximum. 
Owing to the high degree of linear polarization and the high signal-to-noise ratio of the individual pulses, the continuous cycling of the position angle over this range of pulse longitudes can also be clearly seen in the animations provided as supplementary material\footnote{\url{https://www.dropbox.com/sh/1dbb3wnfrx1tt9j/AAB2hONBxf8ChwcdSleDJrJIa?dl=0}}.

\section{Discussion} \label{sec:discussion}

In light of the complexities outlined in the introduction, it would be premature to interpret our results in terms of magnetospheric emission physics without a more comprehensive overview of single-pulse polarimetry across the radio spectrum.  Instead, we aim to add valuable pieces to the puzzle and focus our discussion on two key features of the polarization of the drifting sub-pulses: 1) the dramatic change in the modulation-resolved polarization pattern seen in \Fig{P3fold} that coincides with the longitudinal transition between orthogonally polarised modes near $\phi\simeq\opmswitch$; and 2) the toroidal distributions and cyclic drifting of polarization state around an elliptical path as a function of modulation phase, observed over the region $\partorusleft\lesssim\phi\lesssim\partorusright$.

First, the change in $P_3$-modulated polarization near a pulse longitude of $\phi\sim\opmswitch$ can be explained as a transition between primarily incoherent ($\phi\lesssim\opmswitch$) and partially coherent ($\phi\gtrsim\opmswitch$) mode superposition.
At $\phi\lesssim\opmswitch$, the distribution of polarization states in the Poincar\'e sphere has, to first order, a prolate spheroidal shape as expected for the incoherent superposition of OPMs \citep[e.g., Fig.~\ref{fig:75_torus_1919_1}][]{vt17}.
Over the region roughly defined by $\cmaxleft\lesssim\phi\lesssim\cmaxright$, the polarization state varies as a function of modulation phase between a low degree of polarization (both linear and circular), as expected for an incoherent superposition of OPMs, and a high degree of circular polarization, as expected
for a coherent superposition of linearly polarised natural modes in a highly relativistic plasma \citep[e.g.][]{km98}.
Further evidence of coherent superposition is displayed on either side of this region. Over $\opmswitch\lesssim \phi\lesssim\ycohright$, the position angle varies between approximately $-60\deg$ and $75\deg$; and over $\cmaxright\lesssim\phi\lesssim\partorusleft$, the position angle varies between approximately $30\deg$ and $75\deg$.  These approximately $\pm 45\deg$ offsets can also be interpreted as transitions between incoherent OPM superposition, dominated by one mode, and a coherent superposition of linearly polarised modes \citep{dyk19}.  The sign of the position angle offset is opposite on either side of the region where the degree of circular polarization reaches its maximum, which can be interpreted as monotonic variation of the phase of the coherent superposition of linearly polarised natural modes as a function of drift phase (and hence pulse longitude), as also considered for PSR~B1451$-$68 \citep{dwi21}.

Over $\torusleft\lesssim\phi\lesssim\torusright$, the single-pulse polarization state exhibits remarkable longitude-resolved toroidal distributions that are clear in both the integrated scatter plot shown in \Fig{donut_no_norm} and the normalised scatter plots shown in Figures~\ref{fig:torus_1919_1} and \ref{fig:94_torus_1919_1}.
Furthermore, the polarization state cycles around the elliptical path of the torus in a continuous periodic manner, rather than a stochastic one, modulated in phase with the drifting sub-pulse pattern observed in the total intensity.
Cycling of the polarization state around an elliptical path
is demonstrated by the cross-correlations of Stokes Q and U in \Fig{XcorrPhBin} and in the $P_3$-folded position angle in \Fig{P3fold}. 
The longitude-resolved cross-correlations between the Stokes parameters also indicate that the axis of elliptical revolution precesses about the Stokes V axis as a function of pulse longitude.

To date, continuous cyclic variation of the Stokes polarization vector around an elliptical locus in Poincar\'e space has been observed in only three other sources, PSR~B0320+39, PSR B0809+74, and PSR~B0818$-$13 (observed at $\sim$ 360 MHz, 1380 MHz, and 330  MHz, respectively; E04).  
For PSR~B0320+39, E04 reports significant periodic fluctuations of the polarization vector about an elliptical locus over a range of pulse longitude ($\sim 18\deg$ to $\sim 19\deg$ in Figure 4 of E04) where periodic modulations of the total intensity are significantly diminished. 
In our analysis of PSR~B1919+21, barely detectable modulation of total intensity is observed in the ``saddle region'' ($\saddleleft\lesssim\phi\lesssim\saddleright$; cf. panels c) and d) of \Fig{stat_verif_1919_1} just prior to the range of pulse longitudes where the toroidal distributions are observed ($\torusleft\lesssim\phi\lesssim\torusright$).
As noted by \cite{bac73}, the decreased intensity modulation in this saddle region could be due to overlapping sub-pulses from the drift band regions on either side, especially if they differ by $180\deg$ in modulation phase.
For PSR~B0818$-$13, the elliptical locus reported by E04 appears to occur in a plane that is perpendicular to the Stokes Q--U plane and above it, as inferred from the positive ellipticity angles at pulse longitudes between $\sim 38\deg$ and $\sim 40\deg$ in Figure 2 of E04 and the meridional distribution of polarization states in Figure 5 of E04.
This is quite different to the toroidal distribution of polarization states observed in our analysis of PSR~B1919+21, for which the angle between the axis of revolution and the Stokes V axis varies between $\sim 15\deg$ and $\sim 45\deg$ (cf. Fig. \ref{fig:94_torus_1919_1}).

\subsection{Origin of the toroidal distribution of polarization state}

We consider two plausible explanations for the observed toroidal distributions of polarization state.  These include
1) generalised Faraday rotation; and 2) four-mode mixing.

\textit{Generalised Faraday rotation $-$}
As in previous studies \citep[e.g.][]{cr79,km98,es04,jon16a,dyk17b}, consider a coherent superposition of orthogonally polarised {\it natural} modes, which have anti-parallel Stokes polarization vectors that define an axis in Poincar\'e space.
Phase-coherent superposition of the natural modes produces a new polarization state that lies in a plane that is normal to this axis.
The azimuthal direction of the new polarization state in this normal plane depends on the phase relation between the orthogonal modes.
(In the special case of Faraday rotation, the natural modes have left-handed and right-handed circular polarization, and birefringence-induced differential phase between these modes rotates the polarization vector about the Stokes V axis.)
If the phase difference between the modes varies smoothly between 0 and $2\pi$ during one drift cycle, then the polarization state 
sweeps through a circle in the normal plane, creating a toroidal distribution.

\textit{Four-Mode Mixing $-$} Consider the incoherent superposition of polarised radiation from two emission regions, $\mathcal{M}_1$ and $\mathcal{M}_2$. The emission from each region oscillates periodically between two orthogonally polarised states: $M_1$ and $-M_1$ from region $\mathcal{M}_1$, and $M_2$ and $-M_2$ from region $\mathcal{M}_2$, such that the radiation consists of an incoherent superposition of four modes in total \citep[in contrast to the \emph{coherent} superposition of four modes considered by][]{dwi21}. The primary axes defined by these pairs of OPMs in the Poincar\'e sphere are offset by some angle close to 90$\deg$ (e.g.\ the emission from region $\mathcal{M}_1$ could oscillate between positive and negative Stokes Q and the emission from region $\mathcal{M}_2$ could oscillate between positive and negative Stokes U).
On its own, amplitude modulation of a single pair of incoherently superposed OPMs would create a prolate distribution of polarization states with its primary axis aligned with the axis defined by the OPMs in Poincar\'e space.
If the two pairs of OPM oscillations are  $90\deg$ out of phase with respect to each other, then their incoherent superposition will generate a toroidal distribution of polarization vectors in the Poincar\'e sphere (as depicted in \Fig{cartoon}). 

\subsection{Origin of drifting sub-pulses}

To explain the drifting of the observed sub-beams, we consider two physical models. These include 1) the refractive steering model; and 2) the rotating carousel model.

\textit{Refractive Steering Model $-$} In this model, sub-beams are steered across the primary beam owing to refraction in the pulsar magnetosphere \citep[e.g.][]{ba86,lp98,pl00}.
Each sub-beam consists of a coherent superposition of natural modes and, as the beam is steered, the relative phase difference between the modes slowly varies, causing the resulting polarization state to move in an arc around the primary axis defined by the OPM pair. Because the magnetospheric conditions that determine the refracted ray path of each sub-beam \citep[e.g.][]{fl04,pet00} may also determine the polarization state \citep[e.g., through birefringent effects][]{cr79,lp99}, it seems plausible that periodic modulation of these magnetospheric conditions would result in temporal correlation between the angle at which the refracted sub-beam leaves the pulsar magnetosphere and the polarization state of the sub-beam.
However, more detailed modeling is required to explain how the relative phase between the natural modes cycles through approximately $2\pi$ per drift cycle. 
Furthermore, at pulse longitudes $\torusleft\lesssim\phi\lesssim\torusright$, the polarization state rotates in a plane that is offset by only a small angle from the Q--U plane.  This seems to suggest that the natural modes of the magnetospheric plasma that is responsible for the birefringence are highly circularly polarised, which is inconsistent with the properties of a highly relativistic plasma for which the natural modes are expected to be linearly polarised \cite[e.g.][]{km98}.  
This inconsistency could be resolved by introducing two different birefractive regions in the pulsar magnetosphere, both with linearly polarised natural modes, such that the first region introduces the torus and the second region rotates the primary axis of the torus out of the Q--U plane.

\textit{Rotating Carousel Model $-$} In the rotating carousel model, charged particle streams circulate around the magnetic poles owing to \textbf{E}$\times$\textbf{B} drift, causing sub-beams to drift across the primary beam \citep[e.g.][]{rs75,ran93,rrl+06}.
\citet{rr03} extend this model such that each sub-beam is split (e.g.\ owing to birefringence) into two spatially separated beams; as our line of sight passes from one sub-beam to the next, the polarization state switches between OPMs.
For elliptical cycling of polarization state to arise requires a fortuitous incoherent superposition of two distinct rotating carousels, one carousel comprised of the OPM pair from region $\mathcal{M}_1$ and the other comprised of the OPMs from region $\mathcal{M}_2$ (as described in the four mode mixing model).
The sub-beams in the $\mathcal{M}_1$ carousel must also be offset in magnetic longitude from the sub-beams in the $\mathcal{M}_2$ carousel, such that the polarization state periodically cycles through the four modes and completes a full $360\deg$ rotation with each drift cycle.
It is important to note that, if the sub-beams are as well resolved as shown in Figure~6 of \citet{rr03}, then the total intensity would cycle through four maxima in the time required for the polarization state to complete one circle.
This would not be consistent with our observations; e.g.\ Figure~\ref{fig:2DAC_b} shows that the total intensity is only weakly modulated compared to the much deeper modulations of Stokes Q, U, and V.
This can be explained if the sub-beams observed in the toroidal region are unresolved (i.e.\ the width of each sub-beam is comparable to the spacing between them) such that any remaining modulation in intensity arises from the (possibly small) differences in the peak intensities of the beams along the line of sight that cuts through them (as depicted in \Fig{cartoon}).

In principle, four-mode mixing by a pair of unresolved or partially resolved carousels is consistent with either the barely resolved conal quadruple \citep{omr19} or the five-component multiple \citep{rsw89} classifications that have been applied to describe the mean profile of \psr.
In this interpretation, the first and third drift-bands represent the leading and trailing edges of the outer cone, the second drift-band represents the leading edge of the inner cone, the trailing edge of the inner cone is unresolved, and the inner and outer cones represent the $\mathcal{M}_1$ and $\mathcal{M}_2$ carousels of orthogonally polarised sub-beams.
As noted in the introduction, quantitatively testing this interpretation of the beam geometry is challenging.

First, it is not possible to model the entire P.A.\ curve, plotted in panel b) of \Fig{stat_verif_1919_1}, using the RVM.
Therefore, to apply the RVM it is necessary to apply judgement and model only limited regions of pulse longitude \cite[e.g.][]{ew01}. 
We identify two candidate regions: A) the steepest, negative slope of the P.A.\ profile ranging from $-5\deg$ to $-2.5\deg$ in pulse longitude; and B) the shallow, positive slope from $-13\deg$ to $+2\deg$, excluding $-7.5\deg$ to $-2\deg$. 
Region A exhibits the highest degree of circular polarization and the P.A.\ in this region most likely represents a distortion produced by coherent superposition of natural modes rather than a regular RVM swing \citep{dyk19}.
Moreover, assuming that the leading component centred at $-12\deg$ can be neglected as anomalous, the steepest slope of P.A.\ swing in region A leads the centre of the total intensity profile, which conflicts with the expected effects of aberration and retardation \citep{bcw91}.
%
%
Regardless of the selected region, the best-fit value for the colatitude of the magnetic axis $\alpha$ approaches $0\deg$, which is inconsistent with the  observed narrow pulse width.
Therefore, we abandon any further interpretation of the beam geometry based on the P.A.\ profile.

%
%
%
%
%

To measure the apparent longitudinal offset between the centres of the putative inner and outer cones of emission, we model the mean total intensity pulse profile as a sum of six components, each described by a von Mises function \citep[e.g.][]{wj08}.
The derived difference of $0.36 \deg$ in relativistic phase shifts corresponds to an offset in the inner and outer cone emission heights of $\sim 200$~km \citep{drh04}; therefore, relativistic aberration is a plausible explanation for the overlapping trailing components of the inner and outer cones.

\section{Conclusions} \label{sec:conclusions}

The brightness of PSR~B1919+21, combined with the sensitivity of Arecibo,  enables a variety of approaches to studying the polarization of its drifting sub-pulses.
The methods adopted in this study facilitated the discovery of toroidal and partially toroidal distributions of the Stokes parameters. 
Further investigation showed that, over the range of pulse longitudes that exhibit toroidal distributions, the polarization state rotates continuously and synchronously with the sub-pulse drift period.

This analysis is one step forward in the development of a framework that can be used to interpret the fourth-order statistics of the Stokes parameters in weaker sources.
Eigenvalues can be used to identify rare oblate distributions of the Stokes polarization vectors,
and the cross-correlations between the Stokes parameters can reveal the continuous cycling of polarization state around an elliptical path, synchronous with the intensity modulation of drifting sub-pulses.
In the Arecibo observations of PSR~B1919+21, these interpretations are supported by visualization of the single pulses using methods that are applicable only at high $S/N$. For example, the scatter plots of the normalised Stokes parameters clearly show the toroidal distribution of polarization states and the $P_3$-folded position angle confirms the continuous cycling of polarization state.

 Our study demonstrates that novel statistical methods can facilitate the discovery of relatively rare phenomena, even when analysing decades-old archival observations of the most studied pulsars. Although only 6 pulsars are currently known to exhibit drift-synchronous modulation of polarization state, we speculate that more such sources remain to be discovered.
To obtain a large sample of uniformly studied pulsars, we are conducting the Pulsar Radio Emission Statistics Survey (PRESS)\footnote{\url{https://sites.google.com/view/psr-press}}, an observing campaign that will record single-pulse observations of around 200 pulsars using the ultra-wide bandwidth low-frequency (UWL) receiver at the Parkes Observatory \citep{hmd+20}. 
These data will be made public as soon as they are calibrated and various quality assurance checks have been performed.
All of the software required to perform the analysis presented in this paper is freely available as part of {\sc psrchive}, an open-source C++ development library for the analysis of pulsar astronomical data \citep{hvm04,vdo12}.

\begin{acknowledgements}
The authors dedicate this paper to Professor Jocelyn Bell-Burnell, who discovered the first pulsar, PSR~B1919+21, in 1967. The authors thank Joanna Rankin for providing the software required to interpret the public data of \citet{hr10}; this software is now incorporated into {\sc psrchive}.
We also thank Joanna Rankin for providing valuable comments that greatly improved this paper.
C. Tiburzi is supported by Veni grant (project number 016.Veni.192.086) awarded by the Dutch Research Council (NWO). J. Dyks was supported by grant 2017/25/B/ST9/00385 of National Science Centre, Poland.
Figures 5 through 8 of this paper use the perceptually uniform colour maps developed by \citet{kovesi2015good}\footnote{\url{https://colorcet.com/}}.
At the time of recording the observations presented in this paper, the Arecibo Observatory was part of the National Astronomy and Ionosphere Center, operated by Cornell University for the National Science Foundation. The Arecibo Observatory is currently a facility of the National Science Foundation operated under cooperative agreement by the University of Central Florida and in alliance with Universidad Ana G. Mendez, and Yang Enterprises, Inc. We regret that, while drafting this manuscript, the Arecibo radio telescope's instrument platform crashed into its primary reflector after the cables supporting the platform catastrophically snapped.
\end{acknowledgements}

\bibliographystyle{aa}
\bibliography{psrrefs/journals,psrrefs/psrrefs,psrrefs/modrefs,local}

\begin{appendix} 

\section{Additional Figures}

\begin{figure}[hp!]
\begin{tabular}{c|c}
\includegraphics[width=86mm,trim=42mm 0 30mm -3mm,clip]{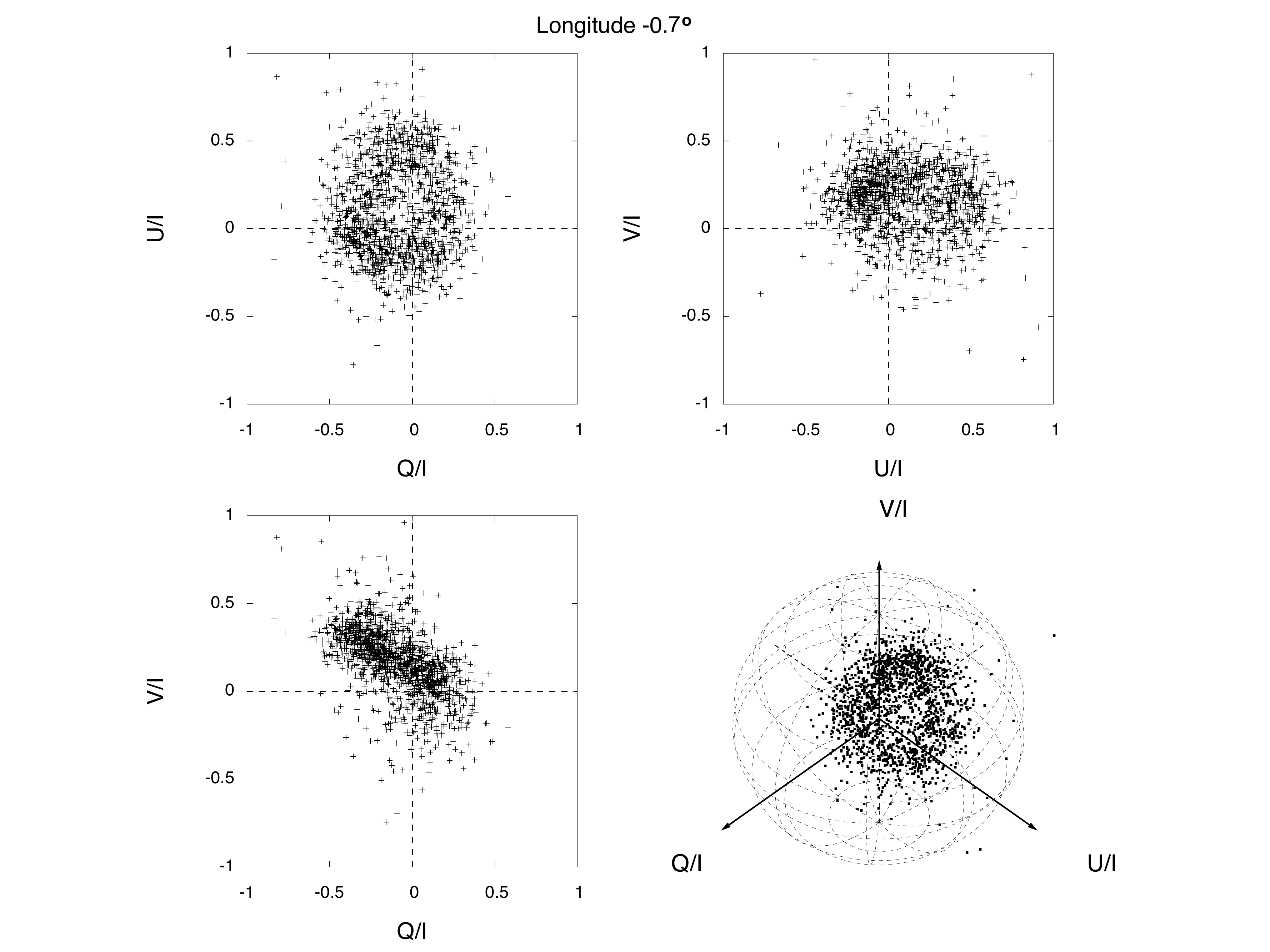} &
\includegraphics[width=86mm,trim=42mm 0 30mm -3mm,clip]{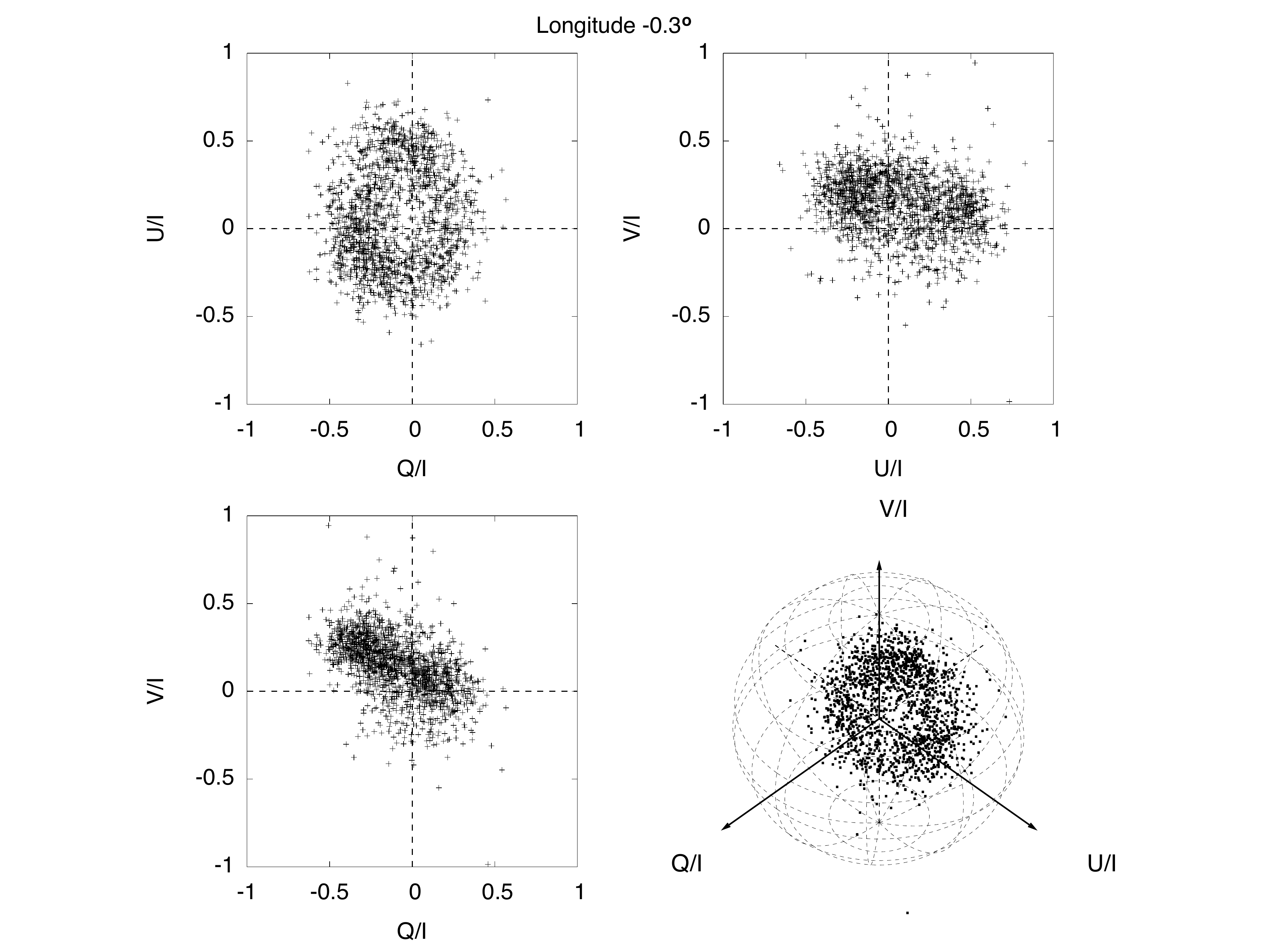} \\
\hline
\includegraphics[width=86mm,trim=42mm 0 30mm -3mm,clip]{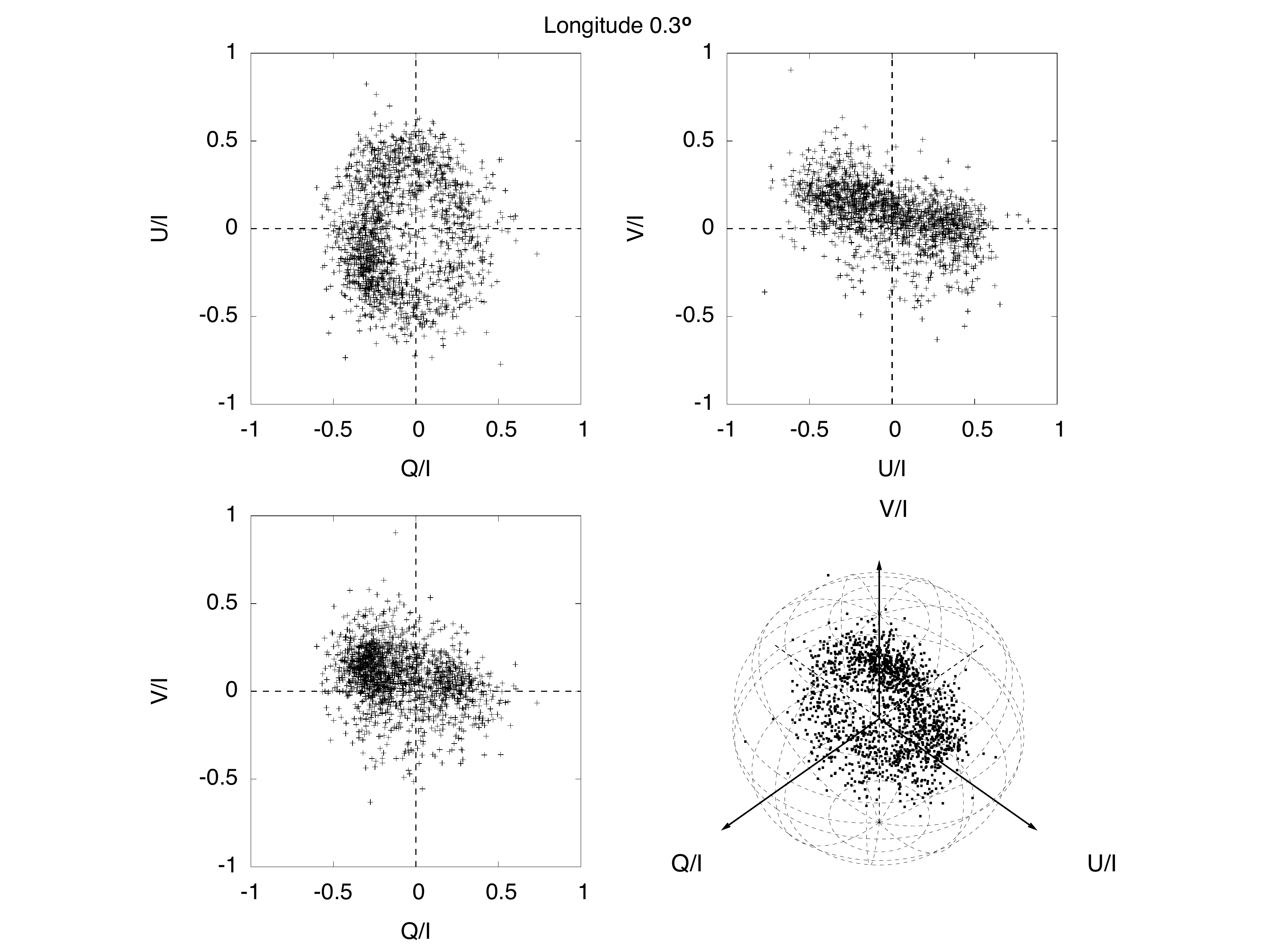} &
\includegraphics[width=86mm,trim=42mm 0 30mm -3mm,clip]{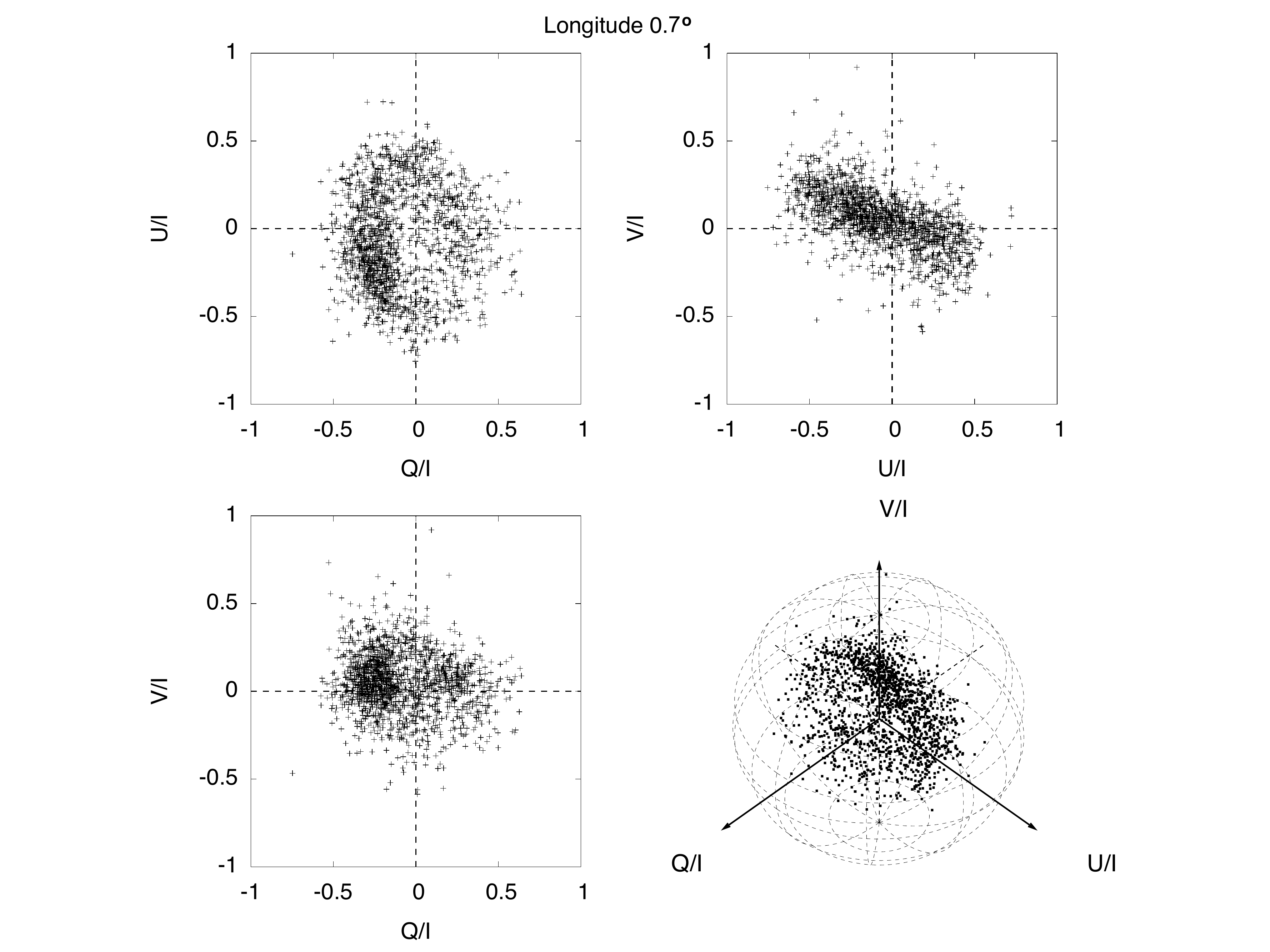}
\end{tabular}
\caption{Distributions of the normalised Stokes parameters in the Poincar\'e sphere at pulse longitudes of approximately $-0.7\deg$ (upper left), $-0.3\deg$ (upper right), $0.3\deg$ (lower left), and $0.7\deg$ (lower right).  In each quadrant, the two upper panels and the lower-left panel show the projected, 2D distribution of the normalised Stokes parameters in the major planes of Poincar\'e space, and the 3D distribution is displayed in the lower-right panel.}
\label{fig:94_torus_1919_1}
\end{figure}

\newpage

\begin{figure*}
\centering
\includegraphics[width=86mm,trim=42mm 0 30mm -3mm,clip]{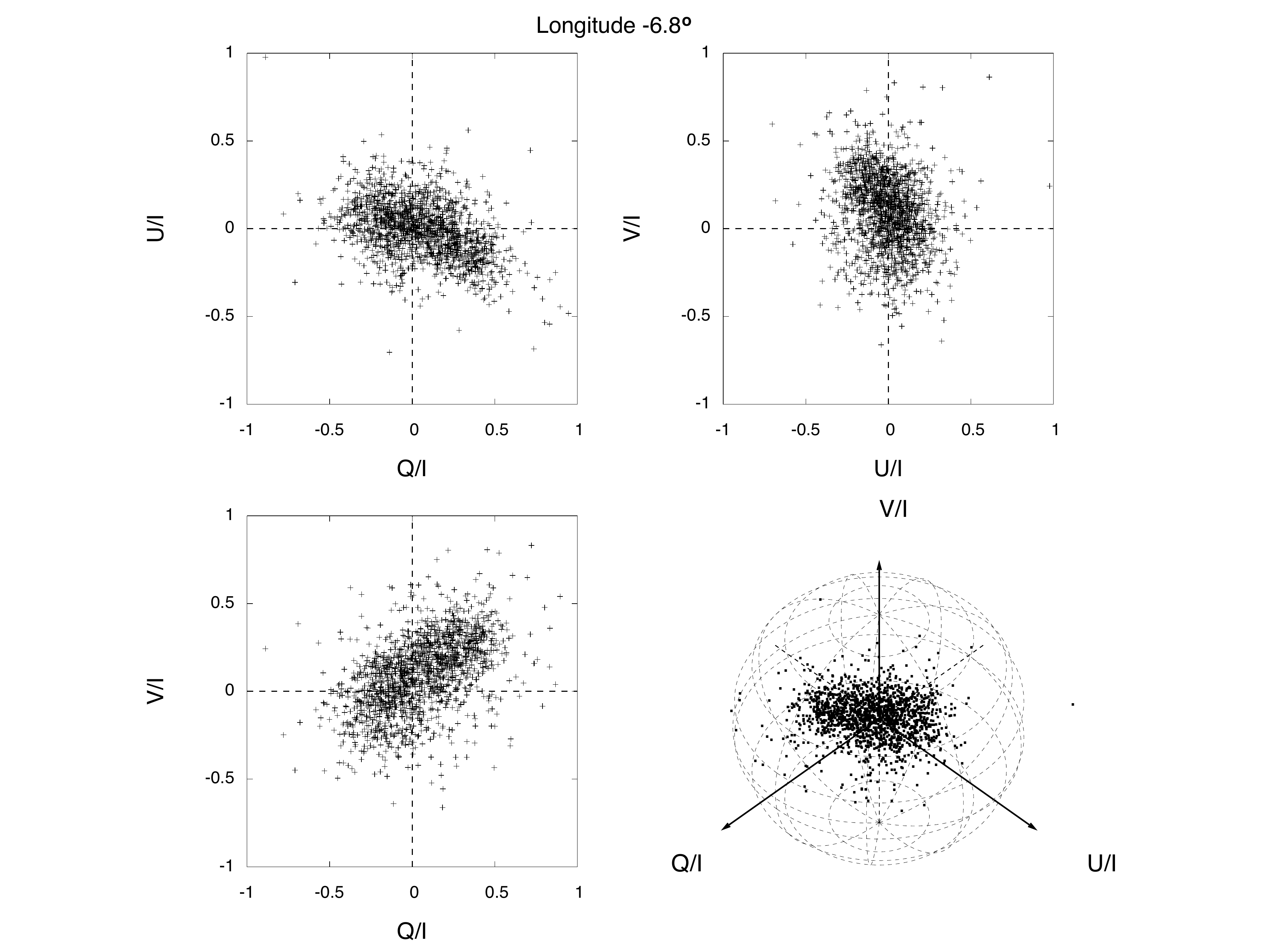}
\caption{Prolate spheroidal distribution of the normalised Stokes parameters in the Poincar\'e sphere at pulse longitude $\sim-$6.8$\deg$. The two upper panels and the lower-left panel show the projected, 2D distribution of the normalised Stokes parameters in the major planes of Poincar\'e space, and the 3D distribution is displayed in the lower-right panel.}
\label{fig:75_torus_1919_1}
\end{figure*}


\begin{figure*}
\centering
\includegraphics[scale=0.35]{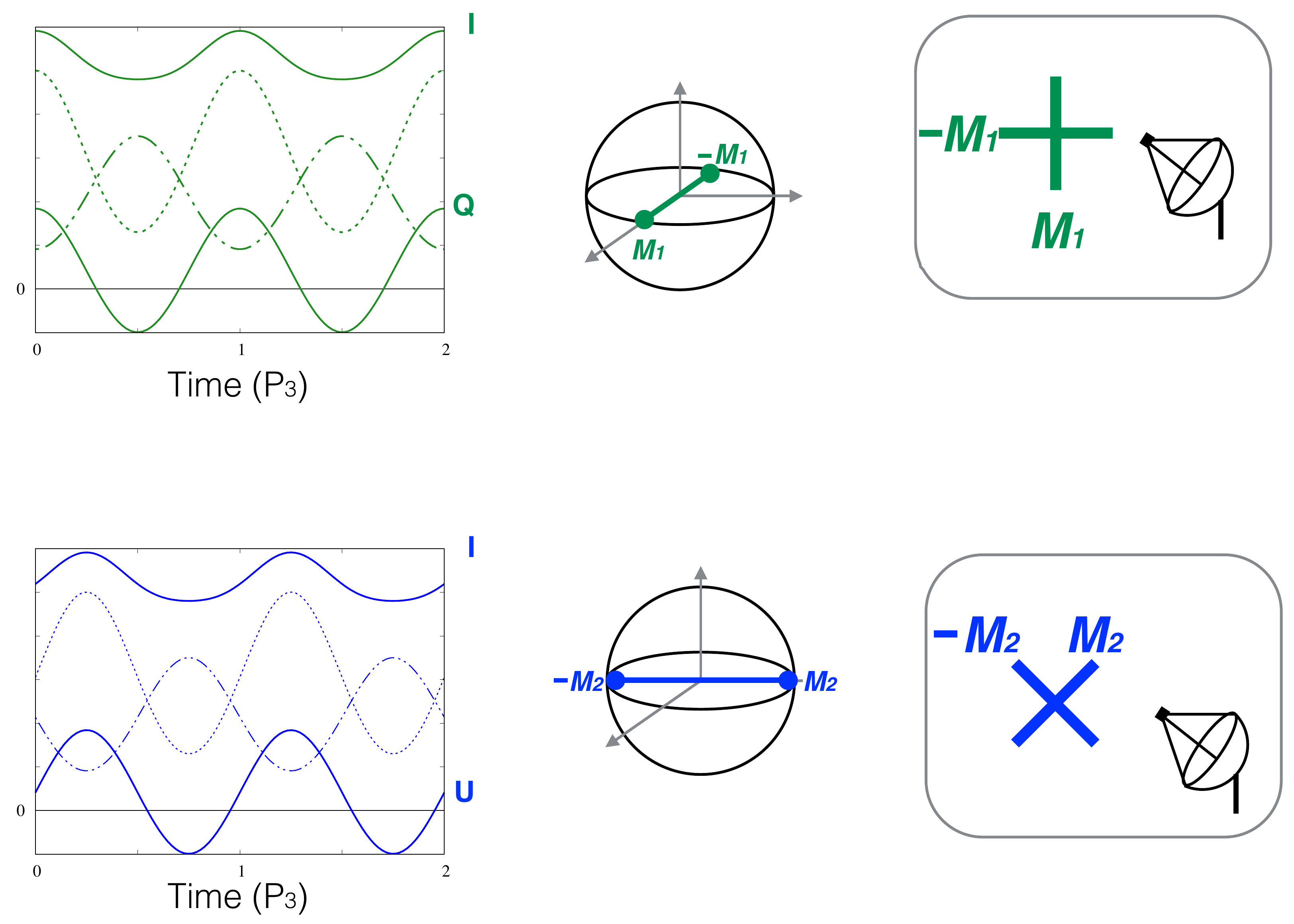}\\
\includegraphics[scale=0.29]{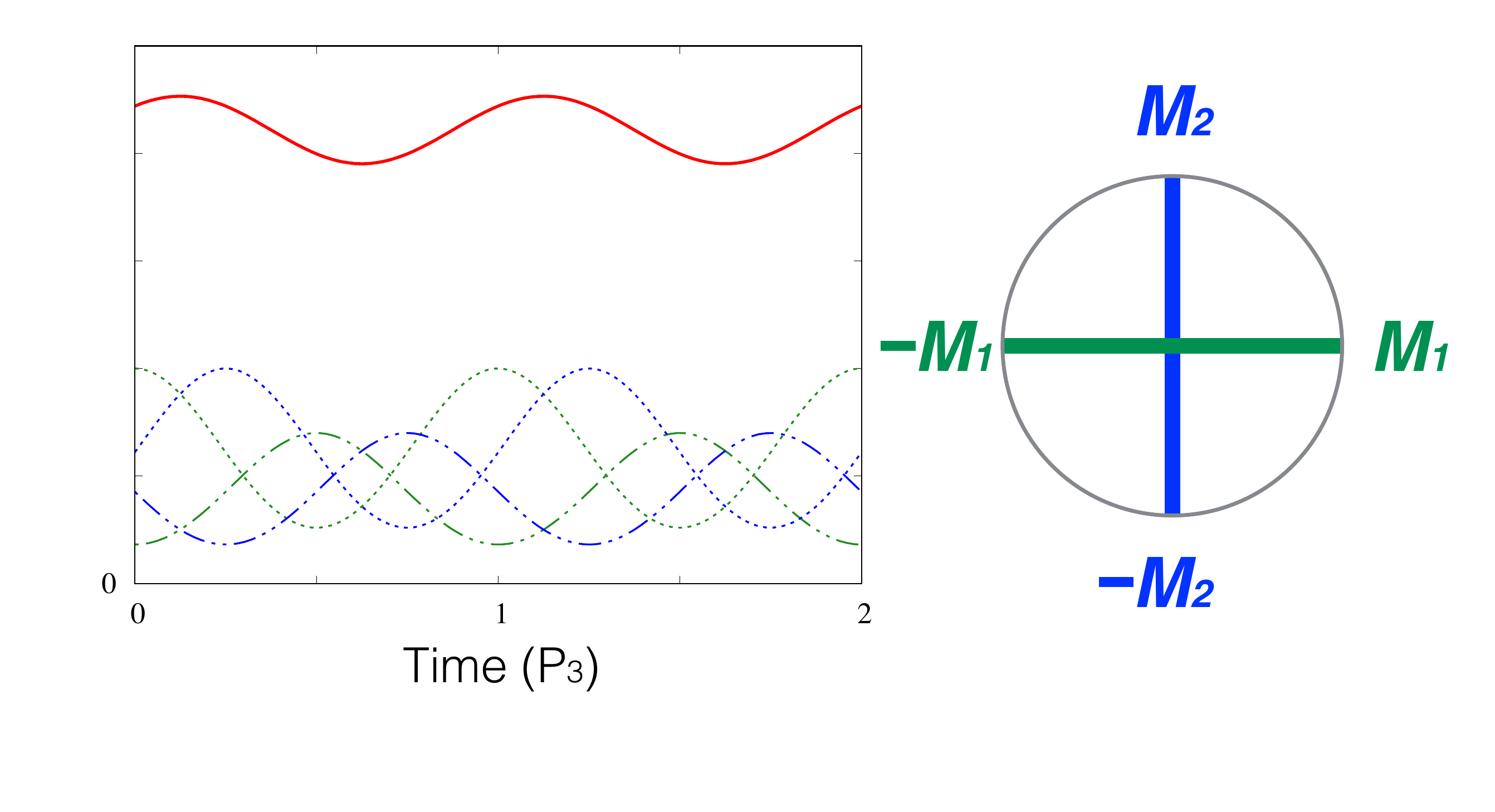}
\caption{Four-mode mixing model. \emph{Top row}: signal from emission region $\mathcal{M}_1$; \emph{middle row}: signal from emission region $\mathcal{M}_2$;
and \emph{bottom row}: incoherent superposition of $\mathcal{M}_1$ and $\mathcal{M}_2$. \emph{Upper two rows}: from right to left, the right-most panels depict the OPM pairs as linearly polarised states in the $x - y$ plane orthogonal to the line of sight of the antenna; the middle panels depict the OPM pairs in the Poincar\'e sphere; and the left-most panels depict the modulated time series produced by the OPM pairs over two drift cycles, where the dashed lines represent the positive OPM state, the dash-dot lines represent the negative OPM state, the upper thick lines represent the total intensity, and the lower thick lines represent the superposed polarization state. 
\emph{Bottom row}: the right panel depicts the four modes in the equatorial plane of the Poincar\'e sphere and the left panel depicts the modulated time series produced by the incoherent superposition the OPM pairs over two drift cycles  (colours and line types are the same as for the upper two rows, and the resultant total intensity is shown in red.}
\label{fig:cartoon}
\end{figure*}

\end{appendix}

\end{document}